\pdfoutput=1
\documentclass{LMCS}

\def\dOi{10(4:19)2014}
\lmcsheading%
{\dOi}
{1--23}
{}
{}
{Jan.~29, 2014}
{Dec.~29, 2014}
{}

\ACMCCS{[{\bf Theory of computation}]: Formal languages and
  automata theory---Automata over infinite objects; Theory and
  algorithms for application domains---Machine learning
  theory---Active learning}

\usepackage{graphicx}
\usepackage{longtable}
\usepackage{amssymb,amsmath}
\usepackage{xspace} 
\usepackage{stmaryrd}
\usepackage[ruled,algo2e,norelsize]{algorithm2e}
\usepackage{mathtools}
\usepackage{hyperref}

\def\N{\ensuremath{\mathbb N}\xspace}
\def\Nz{\ensuremath{\N_{{}>0}}\xspace}
\newcommand\df{\ensuremath{\mathrel{\smash{\stackrel{\scriptscriptstyle{
    \text{def}}}{=}}}}}
\newcommand{\pto}{\rightharpoonup}
\newcommand{\set}[1]{\{1,\ldots,#1\}}
\newcommand\powerset[1]{2^{#1}}
\newcommand\emptyInj{\emptyset}
\newcommand\domain[1]{dom(#1)}

\newcommand{\Data}{D}
\newcommand{\gr}[1]{\textcolor[gray]{0.5}{\scalebox{0.6}{#1}}}
\newcommand{\firstocc}[1]{\mathit{first}_{#1}}
\newcommand{\lastocc}[1]{\mathit{last}_{#1}}

\def\A{\mathcal A}
\def\B{\mathcal B}
\def\cB{\mathcal B}
\newcommand{\Reg}{\mathcal{R}}
\newcommand{\rReg}{\Reg^\uparrow}
\newcommand{\lReg}{\Reg^\odot}
\newcommand{\gReg}{\Reg^\circledast}
\newcommand{\extReg}{\gReg \mathrel{\cup} \lReg \mathrel{\cup} \rReg}
\newcommand{\rreg}[1]{#1^\uparrow}
\newcommand{\lfresh}[1]{#1^\odot}
\newcommand{\gfresh}[1]{#1^\circledast}
\def\Prom{P}
\newcommand{\regof}[1]{\textsf{reg}(#1)}
\newcommand{\opof}[1]{\textsf{op}(#1)}
\newcommand{\Reusable}{\mathit{Free}}
\newcommand{\fStates}{F}
\newcommand{\States}{S}
\newcommand{\init}{\iota}
\newcommand{\Trans}{\Delta}
\newcommand{\xregtrans}[1]{\xrightarrow{#1}}
\newcommand{\conf}{\gamma}
\newcommand{\regm}{\tau}
\newcommand{\Used}{U}
\newcommand{\xconfrel}[1]{\xRightarrow{#1}}
\newcommand{\ack}{\mathsf{ack}}
\newcommand{\req}{\mathsf{req}}
\newcommand{\cOm}{\mathsf{com}}
\newcommand{\forw}{\mathsf{forw}}
\newcommand{\join}{\mathsf{join}}

\def\concretization{\gamma}
\newcommand{\Nat}{\Gamma}
\newcommand{\kNat}[1]{\Gamma_{#1}}
\newcommand{\symbL}{L_\mathit{symb}}
\def\Injection{\mathrm{Inj}}
\newcommand{\can}[1]{#1^{C}}
\newcommand{\snf}{\mathit{snf}}
\newcommand{\DWords}{\gamma}
\newcommand{\DWordsof}[1]{\DWords(#1)}
\newcommand{\WF}{\mathsf{WF}}
\newcommand{\Bounded}[1]{\mathsf{DW}_{#1}}
\newcommand\NF{\mathsf{SNF}}

\newcommand{\mylabel}{\textup{label}}
\renewcommand{\phi}{\varphi}
\newcommand{\ttrue}{\mathit{true}}

\newcommand{\newk}{k}

\newcommand{\param}{\pi}
\newcommand{\runf}{\alpha}
\newcommand{\dataf}{\beta}
\newcommand{\kphi}{k_\dataf}
\newcommand{\tupleI}{\overline{I}}

\def\figurename{Figure}

\usepackage[pdflatex=true,recompilepics=false]{gastex} 
\gasset{frame=false}

\begin{document}

\title[A Robust Class of Data Languages and an Application to Learning]{A Robust Class of Data Languages and an Application to Learning\rsuper*}

\author[B.~Bollig]{Benedikt Bollig\rsuper a}
\address{{\lsuper a}LSV, ENS Cachan, CNRS \& Inria, France}
\email{bollig@lsv.ens-cachan.fr}

\author[P.~Habermehl]{Peter Habermehl\rsuper b}
\address{{\lsuper b}Univ Paris Diderot, Sorbonne Paris Cit\'e, LIAFA, CNRS, France}
\email{haberm@liafa.univ-paris-diderot.fr}

\author[M.~Leucker]{Martin Leucker\rsuper c}
\address{{\lsuper c}ISP, University of L\"ubeck, Germany}
\email{leucker@isp.uni-luebeck.de}

\author[B.~Monmege]{Benjamin Monmege\rsuper d}
\address{{\lsuper d}Universit{\'e} Libre de Bruxelles, Belgium}
\email{bmonmege@ulb.ac.be}

\titlecomment{{\lsuper*}This paper is an extended and revised version of the paper
``A Fresh Approach to Learning Register Automata'' which appeared in DLT 2013}

\begin{abstract}
\noindent
  We introduce \emph{session automata}, an automata model to process data words, i.e., words over an infinite alphabet.
  Session automata support the notion of \emph{fresh} data
  values, which are well suited for modeling protocols in which sessions using
  fresh values are of major interest, like in security protocols or
  ad-hoc networks. Session automata have an expressiveness partly extending, partly reducing that of
  classical register automata. We show that, unlike register automata and their various extensions, session automata are robust: They
  \,(i) are closed under intersection, union, and (resource-sensitive) complementation,
  \,(ii) admit a symbolic regular representation,
  \,(iii) have a decidable inclusion problem
    (unlike register automata), and
  \,(iv) enjoy logical characterizations.
  Using these results, we establish a learning algorithm to infer
  session automata through membership and equivalence queries.
\end{abstract}

\maketitle

\section{Introduction}

The study of automata over data words, i.e., words over an infinite
alphabet, has its origins in the seminal work by Kaminski and Francez
\cite{Kaminski1994}. Their finite-memory automata (more commonly
called \emph{register automata}) equip finite-state machines with
registers in which data values (from the infinite alphabet) can be
stored and be reused later. Register automata preserve some of the
good properties of finite automata: they have a decidable emptiness
problem and are closed under union and intersection. On the other
hand, register automata are neither determinizable nor closed under
complementation, and they have an undecidable equivalence/inclusion
problem. There are actually several variants of register automata,
which all have the same expressive power but differ in the complexity
of decision problems \cite{DL-tocl08,Bjorklund10}.
In the sequel, many more automata models have been introduced (not
necessarily with registers), aiming at a good balance between
expressivity, decidability, and closure properties
\cite{Neven2004,DL-tocl08,KaminskiZ10,BL2010,DBLP:conf/atva/GrumbergKS13,DBLP:conf/lata/GrumbergKS10}.
Some of those models extend register automata, inheriting their
drawbacks such as undecidability of the equivalence problem.

We will follow the work on register automata and study a model that
supports the notion of \emph{freshness}. When reading a data value, it
may enforce that the data value is \emph{fresh}, i.e., it has not
occurred in the whole history of the run. This feature has been
proposed in \cite{DBLP:conf/popl/Tzevelekos11} to model computation
with names in the context of programming-language semantics. Actually,
fresh names are needed to model object creation in object-oriented
languages, and they are
important ingredients in modeling security protocols which often make
use of so-called fresh nonces to achieve their security assertions
\cite{DBLP:conf/ccs/KurtzKW07}. Fresh names are also crucial in the
field of network protocols, and they are one of the key features of
the $\pi$-calculus \cite{MPW92}. Like ordinary register automata,
fresh-register automata preserve some of the good properties of finite
automata. However, they are not closed under complement and also come
with an undecidable equivalence problem.

In this paper, we propose \emph{session automata}, a robust automata
model over data words.
Like register automata, session automata are a syntactical restriction
of fresh-register automata, but in an orthogonal way. Register
automata drop the feature of checking \emph{global freshness} (referring to the
whole history) while keeping a local variant (referring to the
registers). Session automata, on the other hand, discard local
freshness, while keeping the global one. Session automata are
well-suited whenever fresh values are important for a finite period,
for which they will be stored in one of the registers. They correspond
to the model from \cite{BCGK-fossacs12} without stacks. 

Not surprisingly, we will show that session automata and register automata
describe incomparable classes of languages of data words, whereas both
are strictly weaker than fresh-register automata. Contrary to
finite-state unification based automata introduced in~\cite{KamTan06},
session automata (like fresh-register automata) do not have the
capability to reset the content of a register. However, they can test
global freshness which the model of \cite{KamTan06} cannot. The
\emph{variable automata} from \cite{DBLP:conf/lata/GrumbergKS10} do not
employ registers, but rather use bound and free
variables. However, variable automata are close to our model: they use a
finite set of bound variables to track the occurrences of some data values, and a
single free variable for all other data values (that must be different
from data values tracked by bound variables). Contrary to our model,
variable automata cannot test for global freshness, but we are not
able to recognize the language of all data words, contrary to them.

In this paper, we show that session automata
\,(i) are closed under intersection, union, and resource-sensitive
  complementation\footnote{A notion similar to \cite{KST2012}, but for
    a different model.},
\,(ii) have a unique canonical form (analogous to minimal deterministic
  finite automata),
\,(iii) have a decidable equivalence/inclusion problem, and
\,(iv) enjoy logical characterizations.
Altogether, this provides a versatile framework for languages over
infinite alphabets.

In a second part of the paper, we present an application of our
automata model in the area of learning, where decidability of the
equivalence problem is crucial.  Learning automata deals with the
inference of automata based on some partial information, for example
samples, which are words that either belong to the accepted language
or not. A popular framework is that of active learning defined by
Angluin \cite{Angluin:regset} in which a learner may consult a teacher
for so-called membership and equivalence queries to eventually infer
the automaton in question.  Learning automata has many applications in
computer science. Notable examples are the use in model checking
\cite{DBLP:conf/sigsoft/GiannakopoulouM03} and testing
\cite{BergGJLRS05}. See \cite{DBLP:conf/fmco/Leucker07} for an
overview.

While active learning of regular languages is meanwhile well
understood and is supported by freely available libraries such as
LearnLib \cite{MRSL07} and libalf \cite{BKKLNP10}, extensions beyond
plain regular languages are still an area of active
research. Recently, automata dealing with potentially infinite data as
basis objects have been studied. Seminal works in this area are that
of \cite{DBLP:conf/fm/AartsHKOV12,DBLP:conf/sfm/Jonsson11} and
\cite{HowarSJC12}. While the first two use abstraction and refinement
techniques to cope with infinite data, the second approach learns a
sub-class of register automata.  Note that session automata are
incomparable with the model from \cite{HowarSJC12}.  Thanks to their
closure and decidability properties, a conservative extension of
Angluin's classical algorithm will do for their automatic inference.

\subsubsection*{Outline.}
The paper is structured as follows. In Section~\ref{sec:data} we
introduce session automata. Section~\ref{sec:snf-can} presents the
main tool allowing us to establish the results of this paper, namely
the use of data words in symbolic normal form and the construction of
a canonical session automaton. The section also presents some closure
properties of session automata and the decidability of the equivalence
problem. Section~\ref{sec:logic} gives logical characterizations of
our model.  In Section~\ref{sec:learning}, we present an active
learning algorithm for session automata. This paper is an extended
version of \cite{BHLM-dlt2013}.

\section{Data Words and Session Automata}\label{sec:data}

We let $\N$ be the set of natural numbers and $\Nz$ be the set of
non-zero natural numbers. In the following, we fix a non-empty finite
alphabet $\Sigma$ of \emph{labels} and an infinite set $D$ of
\emph{data values}.  In examples, we usually use $D = \N$.  A
\emph{data word} over $\Sigma$ and $\Data$ is a sequence $w =
(a_1,d_1) \cdots (a_n,d_n)$ of pairs $(a_i,d_i) \in \Sigma\times
D$. In other words, $w$ is an element from $(\Sigma\times D)^{*}$. For
$d \in \{d_1,\ldots,d_n\}$, we let $\firstocc{w}(d)$ denote the
position $j \in \{1,\ldots,n\}$ where $d$ occurs for the first time,
i.e., such that $d_j = d$ and there is no $k <j$ such that $d_k =
d$. Accordingly, we define $\lastocc{w}(d)$ to be the last position
where $d$ occurs.

An example data word over $\Sigma = \{a,b\}$ and $D = \N$ is given by
$w=(a,8)(b,4)(a,8)(c,3)$ $(a,4)(b,4)(a,9)$. We have
$\firstocc{w}(4)=2$ and $\lastocc{w}(4)=6$.

This section recalls two existing automata models over data words --
namely register automata, previously introduced in
\cite{Kaminski1994}, and fresh-register automata, introduced in
\cite{DBLP:conf/popl/Tzevelekos11} as a generalization of register
automata. Moreover, we introduce the new model of session automata,
our main object of interest.

Register automata (initially called finite-memory automata) equip finite-state
machines with registers in which data values can be stored and be read
out later. Fresh-register automata additionally come with an oracle that
can determine if a data value is \emph{fresh}, i.e., has not occurred in the
history of a run. Both register and fresh-register automata
are closed under union and intersection, and they have a decidable emptiness
problem. However, they are not closed under complementation, and their
equivalence problem is undecidable, which limits their application in areas such
as model checking and automata learning.
Session automata, on the other hand, are closed under (resource-sensitive)
complementation, and they have a decidable inclusion/equivalence problem.

Given a set $\Reg$, we let $\rReg \df \{\rreg{r} \mid r \in \Reg\}$,
$\lReg \df \{\lfresh{r} \mid r \in \Reg\}$, and $\gReg \df
\{\gfresh{r} \mid r \in \Reg\}$.
In the automata models that we are going to introduce, $\Reg$ will be the set of registers.
Transitions will be labeled with an element
from $\extReg$, which determines a register and the operation that is
performed on it. More precisely, $\gfresh{r}$ writes a globally fresh value into $r$, $\lfresh{r}$
writes a locally fresh value into $r$, and $\rreg{r}$ uses the value that is currently stored in $r$. For $\param\in \extReg$, we let $\regof{\pi} = r$ if
$\pi \in \{\gfresh{r},\lfresh{r},\rreg{r}\}$. Similarly,
\[\opof{\pi}=\begin{cases}
\circledast & \text{if } \param \text{ is of the form } \gfresh{r}\\
\odot & \text{if } \param \text{ is of the form } \lfresh{r}\\
{\uparrow} & \text{if } \param \text{ is of the form } \rreg{r}\,.
\end{cases}\]

\begin{defi}[Fresh-Register Automaton, cf.\ \cite{DBLP:conf/popl/Tzevelekos11}]
  A \emph{fresh-register automaton} (over $\Sigma$ and $\Data$) is a tuple $\A =
  (\States,\Reg,\init,\fStates,\Trans)$ where
  \begin{itemize}
  \item $\States$ is the non-empty finite set of \emph{states},
  \item $\Reg$ is the non-empty finite set of \emph{registers},
  \item $\init \in \States$ is the \emph{initial state},
  \item $\fStates \subseteq \States$ is the set of \emph{final
      states}, and
  \item $\Trans$ is a finite set of \emph{transitions}: each
    transition is a tuple of the form $(s,(a,\param),s')$ where $s,s'
    \in \States$ are the source and target state, respectively, $a \in
    \Sigma$, and $\param \in \extReg$. We call $(a,\param)$ the
    \emph{transition label}.
  \end{itemize}
\end{defi}

\noindent For a transition $(s,(a,\param),s') \in \Delta$, we also write $s
\xregtrans{(a,\param)} s'$. When taking this transition, the automaton
moves from state $s$ to state $s'$ and reads a symbol $(a,d) \in
\Sigma \times D$. If $\param = \rreg{r} \in \rReg$, then $d$ is the
data value that is currently stored in register $r$. If $\param =
\gfresh{r} \in \gReg$, then $d$ is some \emph{globally fresh} data
value, which has not been read in the \emph{whole} history of the run;
$d$ is then written into register $r$. Finally, if $\param =
\lfresh{r} \in \lReg$, then $d$ is some \emph{locally fresh} data
value, which is \emph{currently} not stored in the registers; it will
henceforth be stored in register $r$.

Let us formally define the semantics of $\A$. A \emph{configuration} is a
triple $\conf = (s,\regm,\Used)$ where $s \in \States$ is the current
state, $\regm: \Reg \pto \Data$ is a partial mapping encoding the
current register assignment, and $\Used \subseteq \Data$ is the set of data values that have
been used so far. By $\domain \regm$, we denote the set of registers $r$ such that $\regm(r)$ is defined.
Moreover, $\regm(\Reg) \df \{\regm(r) \mid r \in \domain{\regm}\}$.
We say that $\conf$ is \emph{final} if $s \in
\fStates$. As usual, we define a transition relation over
configurations and let $(s,\regm,\Used) \xconfrel{(a,d)}
(s',\regm',\Used')$, where $(a,d) \in \Sigma \times \Data$, if there is
a transition $s \xregtrans{(a,\param)} s'$ such that the following
conditions hold:
\begin{enumerate}
\item $\begin{cases}
    d = \regm(\regof\param) & \text{if } \opof\param = {\uparrow}\\
    d \not\in \regm(\Reg) & \text{if } \opof\param = \odot\\
    d \not\in \Used & \text{if } \opof\param = \circledast\,,
  \end{cases}$
\item $\domain{\regm'} = \domain{\regm}\cup \{\regof\param\}$ and
  $\Used' = \Used \cup \{d\}$,
\item $\regm'(\regof\param) = d$ and $\regm'(r)=\regm(r)$ for all
  $r\in\domain{\regm}\setminus\{\regof\param\}$.
\end{enumerate}
A run of $\A$ on a data word $(a_1,d_1) \cdots (a_n,d_n) \in (\Sigma
\times \Data)^\ast$ is a sequence
\[\conf_0 \xconfrel{(a_1,d_1)} \conf_1 \xconfrel{(a_2,d_2)} \cdots
\xconfrel{(a_n,d_n)} \conf_n\] for suitable configurations
$\conf_0,\ldots,\conf_n$ with $\conf_0 = (\init,\emptyset,\emptyset)$
(here the partial mapping~$\emptyset$ represents the mapping with
empty domain).
The run is \emph{accepting} if $\conf_n$ is a final configuration. The
\emph{language} $L(\A) \subseteq (\Sigma \times \Data)^\ast$ of $\A$
is then defined as the set of data words for which there is an
accepting run.

Note that fresh-register automata cannot distinguish
between data words that are equivalent up to permutation of data
values. More precisely, given $w,w' \in (\Sigma \times \Data)^\ast$, we write $w\approx w'$
if $w=(a_1,d_1)\cdots (a_n,d_n)$ and $w'=(a_1,d_1')\cdots
(a_n,d_n')$ such that, for all $i,j \in \set{n}$, we have $d_i = d_j$ iff\ $d_i' = d_j'$.
For instance, $(a,4)(b,2)(b,4) \approx
(a,2)(b,5)(b,2)$.
In the following, the equivalence class of a data word
$w$ wrt.\ $\approx$ is written $[w]_{\approx}$.
We call $L \subseteq (\Sigma \times \Data)^\ast$ a \emph{data language} if,
for all $w,w' \in (\Sigma \times \Data)^\ast$ such that $w\approx w'$, we have $w \in L$ if, and only if, $w' \in L$. In particular, $L(\A)$ is a data language for every fresh-register automaton $\A$.

We obtain natural subclasses of fresh-register automata when we
restrict the transition labels $(a,\param) \in \Sigma \times
(\extReg)$ in the transitions.

\begin{defi}[Register Automaton, \cite{Kaminski1994}]
  A \emph{register automaton} is a fresh-register automaton where
  every transition label is from $\Sigma \times (\lReg \cup \rReg)$.
\end{defi}

Like register automata, session automata are a syntactical restriction
of fresh-register automata, but in an orthogonal way. Instead of local
freshness, they include the feature of global freshness.

\begin{defi}[Session Automaton]
  A \emph{session automaton} is a fresh-register automaton where every
  transition label is from $\Sigma \times (\gReg \cup \rReg)$.
\end{defi}

We first compare the three models of automata introduced above in
terms of expressive power.

\begin{exa}\label{ex:req-ack}
  Consider the set of labels $\Sigma = \{\req,\ack\}$ and the set of
  data values $\Data = \N$, representing an infinite supply of process
  identifiers (pids). We model a simple (sequential) system where
  processes can approach a server and make a request, indicated by
  $\req$, and where the server can acknowledge these requests,
  indicated by $\ack$. More precisely, $(\req,p) \in \Sigma \times
  \Data$ means that the process with pid $p$ performs a request, which is
  acknowledged when the system executes $(\ack,p)$.

  \figurename~\ref{fig:multiples3}(a) depicts a register
  automaton that recognizes the language $L_1$ of data words verifying
  the following conditions:
  \begin{itemize}
  \item there are at most two open requests at a time;
  \item a process waits for an acknowledgment before making another request;
  \item every acknowledgment is preceded by a request;
  \item requests are acknowledged in the order they are received.
  \end{itemize}
  In the figure, an edge label of the form $(\req,\lfresh{r_i} \vee
  \rreg{r_i})$ shall denote that there are two transitions, one
  labeled with $(\req,\lfresh{r_i})$, and one labeled with
  $(\req,\rreg{r_i})$. Whereas a transition labeled with
  $(\req,\lfresh{r_i})$ is taken when the current data value does not
  appear currently in the registers (but could have appeared before in
  the data word) and store it in $r_i$, transition labeled with
  $(\req,\rreg{r_i})$ simply checks that the current data is stored in
  register $r_i$. The automaton models a server that can store two
  requests at a time and will acknowledge them in the order they are
  received. For example, it accepts
  $(\req,8)(\req,4)(\ack,8)(\req,3)(\ack,4)(\req,8)(\ack,3)(\ack,8)$.

  When we want to guarantee that, in addition, every process makes at most one request, we
  need the global freshness
  operator. \figurename~\ref{fig:multiples3}(b)
  hence depicts a session automaton recognizing the language $L_2$ of
  all the data words of $L_1$ in which every process makes at most one
  request. Notice that the transition from $s_0$ to $s_1$ is now
  labeled with $(\req,\gfresh{r_1})$, so that this transition can only
  be taken in case the current data value has never been seen
  before. We obtain $\A_2$ from $\A_1$ by replacing every occurrence
  of $\lfresh{r_i}\lor \rreg{r_i}$ with $\gfresh{r_i}$. While
  $(\req,8)(\req,4)(\ack,8)(\req,3)(\ack,4)(\req,8)(\ack,3)(\ack,8)$
  is no longer contained in $L_2$, $(\req,8)$
  $(\req,4)(\ack,8)(\req,3)(\ack,4)(\ack,3)$ is still accepted.

  As a last example, consider the language $L_3$ of data words in
  which every process makes at most one request (without any other
  condition). A fresh-register automaton recognizing it is given in
  \figurename~\ref{fig:fra}.

\begin{gpicture}[name=registerAutomaton,ignore]
    \unitlength=4
    \gasset{Nframe=y,Nw=4,Nh=4,Nmr=4,ilength=3.5,flength=3.5,AHangle=30}
    \node[Nmarks=if,iangle=90,fangle=-90](0)(0,0){$s_0$}
    \gasset{curvedepth=4}%
    \node(1)(-20,0){$s_1$}%
    \node(2)(20,0){$s_2$}%
    \node(12)(0,13){$s_{12}$}%
    \node(21)(0,-13){$s_{21}$}%
          
    \drawedge[curvedepth=-1.5,ELside=r,ELpos=47](0,1){$(\req,\lfresh{r_1} \vee \rreg{r_1})$}
    \drawedge[curvedepth=-1.5,ELside=r](1,0){$(\ack,\rreg{r_1})$}
    \drawedge[ELdist=-0.5](1,12){$(\req,\lfresh{r_2} \vee \rreg{r_2})$}
    \drawedge[ELdist=-1,ELside=l](2,21){$(\req,\lfresh{r_1} \vee \rreg{r_1})$}
    \drawedge[ELdist=0.3](21,1){$(\ack,\rreg{r_2})$}
    \drawedge[ELdist=0.3,ELside=l](12,2){$(\ack,\rreg{r_1})$}
    \drawedge[curvedepth=1.5,ELside=l,ELpos=47](0,2){$(\req,\lfresh{r_2} \vee \rreg{r_2})$}
    \drawedge[curvedepth=1.5,ELside=l](2,0){$(\ack,\rreg{r_2})$}
  \end{gpicture}
  \begin{gpicture}[name=sessionAutomaton,ignore]
    \unitlength=4
    \gasset{Nframe=y,Nw=4,Nh=4,Nmr=4,ilength=3.5,flength=3.5,AHangle=30}
    \node[Nmarks=if,iangle=90,fangle=-90](0)(0,0){$s_0$}
    \gasset{curvedepth=4}%
    \node(1)(-20,0){$s_1$}%
    \node(2)(20,0){$s_2$}%
    \node(12)(0,13){$s_{12}$}%
    \node(21)(0,-13){$s_{21}$}%
          
    \drawedge[curvedepth=-1.5,ELside=r](0,1){$(\req,\gfresh{r_1})$}
    \drawedge[curvedepth=-1.5,ELside=r](1,0){$(\ack,\rreg{r_1})$}
    \drawedge[ELdist=0.3](1,12){$(\req,\gfresh{r_2})$}
    \drawedge[ELdist=0.3,ELside=l](2,21){$(\req,\gfresh{r_1})$}
    \drawedge[ELdist=0.3](21,1){$(\ack,\rreg{r_2})$}
    \drawedge[ELdist=0.3,ELside=l](12,2){$(\ack,\rreg{r_1})$}
    \drawedge[curvedepth=1.5,ELside=l](0,2){$(\req,\gfresh{r_2})$}
    \drawedge[curvedepth=1.5,ELside=l](2,0){$(\ack,\rreg{r_2})$}
  \end{gpicture}
  \begin{figure}[tb]
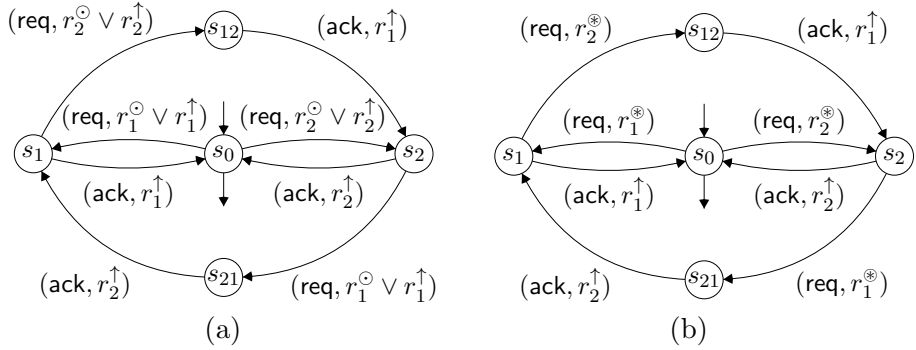

    \centering 
     \begin{tabular}{cc}
      \scalebox{0.9}{\gusepicture{registerAutomaton}} &
      \hspace{1em}\scalebox{0.9}{\gusepicture{sessionAutomaton}}\\
      (a) & (b)
     \end{tabular}
    \caption{(a) Register automaton $\A_1$ for $L_1$, (b) Session
      automaton $\A_2$ for $L_2$}
    \label{fig:multiples3}
  \end{figure}
   \begin{gpicture}[name=freshRegisterAutomaton,ignore]
    \unitlength=4
    \gasset{Nframe=y,Nw=4,Nh=4,Nmr=4,ilength=3.5,flength=3.5,AHangle=30}
    
    \node[Nmarks=if,fangle=-90](s0)(0,0){}
    
    \drawloop[ELside=r,loopCW=n,loopdiam=4,ELdist=-0.5,loopangle=0](s0){
      $\begin{array}{l}
        (\req,\gfresh{r})\\[0.5ex]
        (\ack,\lfresh{r})\\[0.5ex]
        (\ack,\rreg{r})\end{array}$}
  \end{gpicture}
  \begin{figure}[tb]
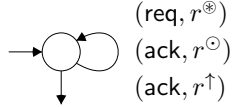

    \centering
    \scalebox{.9}{\gusepicture{freshRegisterAutomaton}}
    \caption{Fresh-register automaton $\A_3$ for $L_3$}
    \label{fig:fra} 
  \end{figure}
\end{exa}

\begin{prop}
  Register automata and session automata are incomparable in terms of expressive power.  Moreover,
  fresh-register automata are strictly more expressive than both
  register automata and session automata.
\end{prop}
\proof
  We use the languages $L_1$, $L_2$, and $L_3$ defined in
  Example~\ref{ex:req-ack} to separate the different automata models.

  First, the language $L_1$, recognizable by a register automaton, is not
  recognized by any session automaton. Indeed, denoting $w_d$ the data
  word $(\req,d)(\ack,d)$, no session automaton using $k$ registers
  can accept
  \[w_1w_2 \cdots w_k w_{k+1} w_k \cdots w_2 w_1\in L_1\,.\]
  Intuitively, the session automaton must store all $k+1$ data values
  of the requests in order to check the acknowledgement, and cannot
  discard any of the $k$ first data values to store the $(k+1)$th
  since all of them have to be reused afterwards (and at that time
  they are not globally fresh anymore). More precisely, after reading
  $w_1w_2\cdots w_k$ the configuration must be of the form
  $(s,\tau,\{1,2,\ldots,k\})$ with $\tau$ being a permutation of
  $\{1,\ldots,k\}$. Reading $w_{k+1}$, with fresh data value $k+1$,
  must then replace the content of a register with $k+1$. Suppose it
  is register $j$. Then, when reading the second occurrence of $w_j$,
  data value $j$ is not globally fresh anymore, yet it is not stored
  anymore in the registers, which does not allow us to accept this
  data word.
  
  Then, the language $L_2$, recognizable by a session automaton, is
  indeed not recognizable by a register automaton, for the same
  reasons as already developed in Proposition~5 of
  \cite{Kaminski1994}. Intuitively, the automaton needs to register
  every data value encountered since it has to ensure the freshness of
  every pid.

  Finally, language $L_3$, recognized by a fresh-register automaton,
  is not recognized by any register automaton (see again Proposition~5
  of \cite{Kaminski1994}) nor by any session automaton. In particular, no
  session automaton with $k$ registers can accept the data word
  \[(\req,1)(\req,2)\cdots (\req,k+1)(\ack,1)(\ack,2)\cdots
  (\ack,k+1)\in L_3\] since when reading the letter $(\req,k+1)$, all
  the $k+1$ data values seen so far should be registered to accept the
  suffix afterwards. A formal proof can be done in the same spirit as
  for $L_1$. \qed

\begin{exa}\label{ex:p2p}
  To conclude the section, we present a session automaton with $2$ registers that models a P2P
  protocol. A user can join a host with address $x$, denoted by action
  $(\join,x)$. The request is either forwarded by $x$ to another host
  $y$, executing $(\forw_1,x)(\forw_2,y)$, or acknowledged by
  $(\ack,x)$. In the latter case, a connection between the user and
  $x$ is established so that they can communicate, indicated by action
  $(\cOm,x)$. Note that the sequence of actions
  $(\forw_1,x)(\forw_2,y)$ should be considered as an encoding of a
  single action $(\forw,x,y)$ and is a way of dealing with actions
  that actually take two or more data values, as considered, e.g., in
  \cite{HowarSJC12}. An example execution of our protocol is
  $(\join,145)(\forw,145,978)(\forw,978,14)(\ack,14)(\cOm,14)(\cOm,14)(\cOm,14)$. In
  \figurename~\ref{fig:P2P}, we show the session automaton for the P2P
  protocol: it uses 2 registers. Following \cite{BCGK-fossacs12}, our
  automata can be easily extended to multi-dimensional data
  words. This also holds for the learning algorithm that will be
  presented in Section~\ref{sec:learning}.

\begin{gpicture}[name=P2P,ignore]
\unitlength=4
\gasset{AHangle=30,ilength=3.5,flength=3.5,Nw=4,Nh=4,Nmr=4}
\node[Nmarks=i](0)(5,0){$0$}
\node(1)(20,0){$1$}
\node(2)(40,0){$2$}
\node[Nmarks=f,fangle=-90](3)(20,-10){$3$}
\node[Nmarks=f,fangle=-90](4)(40,-10){$4$}

\gasset{Nw=5,Nh=5,Nmr=5,loopdiam=4}
\drawedge(0,1){$\join,\gfresh{r_1}$}
\drawedge[curvedepth=1.5](1,2){$\forw,\rreg{r_1},\gfresh{r_2}$}
\drawedge[curvedepth=1.5](2,1){$\forw,\rreg{r_2},\gfresh{r_1}$}
\drawedge[ELside=r](1,3){$\ack,\rreg{r_1}$}
\drawedge[ELside=l](2,4){$\ack,\rreg{r_2}$}
\drawloop[loopCW=n,ELside=r,loopangle=180](3){$\cOm,\rreg{r_1}$}
\drawloop[loopangle=180](4){$\cOm,\rreg{r_2}$}
\end{gpicture}
\begin{figure}[tb]
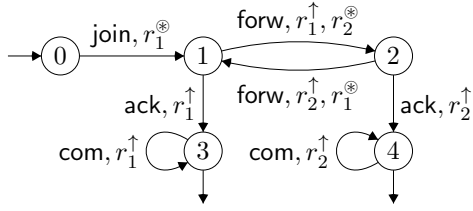

  \centering
  \scalebox{.9}{\gusepicture{P2P}}
  \caption{Session automaton for the P2P protocol}
  \label{fig:P2P}
\end{figure}

\end{exa}

\section{Symbolic Normal Form and Canonical Session Automata}
\label{sec:snf-can}

Closure properties of session automata, decidability of inclusion/equivalence
and the learning algorithm will be established by means of a symbolic
normal form of a data word, as well as a canonical session automaton
recognizing those normal forms. The crucial observation is that data
equality in a data word recognized by a session automaton only depends
on the transition labels that generate it. In this section, we suppose
that the set of registers of a session automaton is of the form $\Reg
= \{1,\ldots,k\}$. In the following, we let $\Nat = \Nz^\circledast
\cup \Nz^\uparrow$ and, for $k \geq 1$, $\kNat{k} = \set{k}^\circledast
\cup \set{k}^\uparrow$.

\subsection{Data Words in Symbolic Normal Forms}

Suppose a session automaton reads a sequence $u = (a_1,\pi_1) \cdots
(a_n,\pi_n) \in (\Sigma \times \Nat)^\ast$ of transition labels. We
call $u$ a \emph{symbolic word}.  It ``produces'' a data word if, and
only if, a register is initialized before it is used. Formally, we say
that $u$ is \emph{well-formed} if, for all positions $j \in \set{n}$
with $\opof{\pi_j} = \mathord{\uparrow}$, there is $i < j$
such that $\pi_i = \gfresh{\regof{\pi_j}}$.  Let $\WF \subseteq
(\Sigma \times \Nat)^\ast$ be the set of all well-formed words.

With $u = (a_1,\pi_1) \cdots
(a_n,\pi_n)\in(\Sigma\times \Nat)^\ast$, we can associate an equivalence
relation $\mathord{\sim}_u$ over $\set{n}$, letting $i \sim_u j$ if,
and only if,
\begin{itemize}
\item $\regof{\pi_i} = \regof{\pi_j}$, and
\item $i\leq j$ and there is no position $k \in
  \{i+1,\ldots,j\}$ such that $\pi_{k} = \gfresh{\regof{\pi_i}}$, or\\
  $j\leq i$ and there is no position $k \in \{j+1,\ldots,i\}$ such
  that $\pi_{k} = \gfresh{\regof{\pi_j}}$.
\end{itemize}
If $u$ is well-formed, then the data values of every data word
$w=(a_1,d_1) \cdots (a_n,d_n)$ that a session automaton ``accepts
via'' $u$ conform with the equivalence relation $\sim_u$, that is, we
have $d_i = d_j$ iff\ $i \sim_u j$.
This motivates the following definition.
Given a well-formed word $u = (a_1,\pi_1) \cdots
(a_n,\pi_n)\in(\Sigma\times \Nat)^\ast$, we call $w \in (\Sigma \times \Data)^\ast$
a \emph{concretization} of $u$ if it is of the form $w=(a_1,d_1) \cdots (a_n,d_n)$
such that, for all $i,j \in \{1,\ldots,n\}$, we have $d_i = d_j$
iff\ $i \sim_u j$. For
example, $w=(a,8)(a,5)(b,8)(a,3)(b,3)$ is a concretization of
$u=(a,\gfresh 1)(a,\gfresh 2)(b,\rreg 1)(a,\gfresh 2)(b,\rreg 2)$.

Let $\DWordsof{u}$ denote the set of all concretizations of
$u$. Observe that, if $w$ is a data word from $\DWordsof u$, then
$\DWordsof u = [w]_{\approx}$. Concretization is extended to sets $L
\subseteq (\Sigma \times \Nat)^\ast$ of well-formed words, and we let
$\DWordsof{L} \df \bigcup_{u \in L \,\cap\, \WF} \DWordsof{u}$. Note
that, here, we first filter the well-formed words before applying the
operator.
Now, let $\A=(\States,\Reg,\init,\fStates,\Trans)$ be a session
automaton. In the obvious way, we may consider $\A$ as a finite
automaton over the finite alphabet $\Sigma \times (\gReg \cup
\rReg)$. We then obtain a regular language $\symbL(\A) \subseteq
(\Sigma \times \Nat)^\ast$ (indeed, $\symbL(\A) \subseteq (\Sigma
\times \kNat k)^\ast$ if $\Reg = \{1,\ldots,k\}$).  It is not
difficult to verify that $L(\A) = \DWordsof{\symbL(\A)}$.

Though we have a symbolic representation of data languages recognized
by session automata, it is in general difficult to compare their
languages, since different symbolic words may give rise to the
same concretizations. For example, we have
$\DWordsof{(a,\gfresh{1})(a,\gfresh{1})(a,\rreg{1})} =
\DWordsof{(a,\gfresh{1})(a,\gfresh{2})(a,\rreg{2})}$. However, we can
associate, with every data word, a symbolic normal form, producing
the same set of concretizations. Intuitively, the normal form uses the
first (according to the natural total order) register whose current
data value \emph{will not be used anymore}. In the above example,
$(a,\gfresh{1})(a,\gfresh{1})(a,\rreg{1})$ would be in symbolic normal
form: the data value stored at the first position in register $1$ is
not reused so that, at the second position, register 1 \emph{must} be
overwritten. For the same reason,
$(a,\gfresh{1})(a,\gfresh{2})(a,\rreg{2})$ is not in symbolic normal
form, in contrast to
$(a,\gfresh{1})(a,\gfresh{2})(a,\rreg{2})(a,\rreg{1})$ where register
$1$ is read at the end of the word.

Formally, given a data word $w=(a_1,d_1) \cdots (a_n,d_n)$, we define
its symbolic normal form $\snf(w) \df (a_1,\pi_1) \cdots (a_n,\pi_n)
\in (\Sigma \times \Nat)^\ast$ inductively, along with sets
$\Reusable(i) \subseteq \Nz$ indicating the registers that are reusable
after executing position $i \in \set{n}$. Setting $\Reusable(0) =
\Nz$, we define
\[\param_{i}=
\begin{cases}
\gfresh{\min(\Reusable(i-1))} & \text{if } i = \firstocc{w}(d_{i})\\
\rreg{\regof{\pi_{\firstocc{w}(d_{i})}}} & \text{otherwise}\,,
\end{cases}\] and
\[\Reusable(i)=
\begin{cases}
  \Reusable(i-1) \setminus \min(\Reusable(i-1)) & \text{if } i =
  \firstocc w(d_i) \neq \lastocc{w}(d_{i})\\
  \Reusable(i-1) \cup \{\regof{\param_i}\} & \text{if } i=\lastocc w(d_i) \\
  \Reusable(i-1) & \text{otherwise}\,.
\end{cases}\] We canonically extend $\snf$ to data languages $L$,
setting $\snf(L) = \{\snf(w) \mid w \,\in\, L\}$.

\begin{exa}
Let $w=(a,8)(b,4)(a,8)(c,3)(a,4)(b,3)(a,9)$.
Then, we have $\snf(w) = (a,\gfresh{1})(b,\gfresh{2})(a,\rreg{1})(c,\gfresh{1})(a,\rreg{2})(b,\rreg{1})(a,\gfresh{1})$.
\end{exa}

\begin{gpicture}[name=dataWordSNF,ignore]
  \gasset{Nframe=y,Nmr=0,Nh=14,AHLength=1.5,AHangle=22}
  \unitlength=1.1mm

  \newcommand{\lettersize}{0.9}

  \node[Nadjust=w](a1)(0,0){$\begin{array}{rl}&w\\[-0.2ex]\in&\DWordsof
      u\\\in&(\Sigma \times \Data)^\ast\\[0.3ex]\end{array}$}
  \node[Nadjust=w](a2)(50,0){$\begin{array}{rl}&u\\[-0.5ex]=&\snf(w)\\\in&(\Sigma
      \times \Nat)^\ast\\[0.3ex]\end{array}$}

  \drawedge[curvedepth=1.5,ELdist=1.5](a1,a2){$\snf$}
  \drawedge[curvedepth=1.5](a2,a1){$\DWords$}

\end{gpicture}

The relation between the mappings $\DWords$ and $\snf$ is illustrated 
below
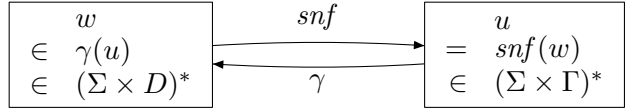
\begin{center}
  \gusepicture{dataWordSNF}
\end{center}
One easily verifies that $L = \DWordsof{\snf(L)}$, for all data
languages $L$.  Therefore, equality of data languages reduces to
equality of their symbolic normal forms:

\begin{lem}\label{lem:dataeq}
  Let $L$ and $L'$ be data languages. Then, $L=L'$ if, and only if,
  $\snf(L) = \snf(L')$.
\end{lem}

Of course, symbolic normal forms may use any number of registers so
that the set of symbolic normal forms is a language over an infinite
alphabet as well. However, given a session automaton $\A$, the
symbolic normal forms that represent the language $L(\A)$ uses only a
bounded (i.e., finite) number of registers. Indeed, an important
notion in the context of session automata is the \emph{bound} of a
data word. Intuitively, the bound of $w = (a_1,d_1) \cdots (a_n,d_n)
\in (\Sigma \times \Data)^\ast$ is the minimal number of registers
that a session automaton needs in order to execute $w$. Or, in other
words, the bound is the maximal number of overlapping \emph{sessions}.
A session is an interval delimiting the occurrence of one particular
data value. Formally, a session of $w$ is a set $I \subset \Nz$ of the
form $\{\firstocc{w}(d),\firstocc{w}(d)+1,\ldots,\lastocc{w}(d)\}$
with $d \in \Data$ a data value appearing in $w$. Given $k
\in\Nz$, we say that $w$ is $k$-\emph{bounded} if every position $i
\in \{1,\ldots,n\}$ is contained in at most $k$ sessions.  Let
$\Bounded{k}$ denote the set of $k$-bounded data words, and let
$\NF_{k} \df \snf(\Bounded{k})$ denote the set of symbolic normal
forms of all $k$-bounded data words.

One can verify that a data word $w$ is $k$-bounded if, and only if,
$\snf(w)$ is a word over the alphabet $\Sigma \times \kNat{k}$. 
Notice that $\Bounded k = \concretization((\Sigma\times \kNat k)^\ast)$.
Indeed, inclusion $\Bounded k \supseteq \concretization((\Sigma\times
\kNat k)^\ast)$ is trivial. If, on the other hand, $w\in\Bounded k$, we must have
$\snf(w)\in (\Sigma\times \kNat k)^\ast$, which implies that
$w\in\concretization(\snf(w))\subseteq \concretization((\Sigma\times
\kNat k)^\ast)$.

A data language $L$ is said to be $k$-\emph{bounded} if $L \subseteq
\Bounded{k}$. It is \emph{bounded} if it is $k$-bounded for some $k$.
Note that the set of all data words is not bounded.

\begin{figure}[t]
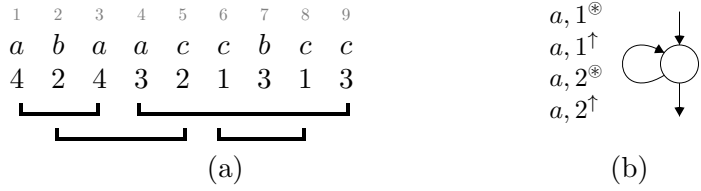

\centering
\begin{tabular}{cc}
\begin{minipage}{.4\linewidth}
  $\begin{array}{ccccccccc}
    \gr{1} & \gr{2} & \gr{3} & \gr{4} & \gr{5} & \gr{6} & \gr{7} & \gr{8} & \gr{9} \\
    a & b & a & a & c & c & b & c & c\\
    4 & 2 & 4 & 3 & 2 & 1 & 3 & 1 & 3
  \end{array}$\\\vspace{-0.8ex}
\hspace*{0.3ex}
  $\underbracket{\hspace*{6.5ex}}$\hspace*{3ex}$\underbracket{\hspace*{17ex}}$\\
\hspace*{4ex}
 $\underbracket{\hspace*{10.5ex}}$\hspace*{2.5ex}$\underbracket{\hspace*{7.1ex}}$
\end{minipage} &
\hspace{2em}
\begin{minipage}{.2\linewidth}
    \raisebox{-2em}{\scalebox{.9}{\gusepicture{simpleSA}}}
    \\\vspace{.3em}
\end{minipage} \\[9mm]
(a) & (b)
\end{tabular}
\caption{(a) A data word and its sessions, (b) Session automaton
  recognizing all 2-bounded data words}
\label{fig:multiples2}
\end{figure}

\figurename~\ref{fig:multiples2}(a) illustrates a
data word $w$ with four different sessions. It is 2-bounded, as no
position shares more than 2 sessions.

\begin{gpicture}[name=simpleSA,ignore]
  \unitlength=4 %
  \gasset{Nw=4,Nh=4,Nmr=4,loopdiam=4}
  \gasset{AHangle=30,ilength=3.5,flength=3.5}
  \node[Nmarks=if,iangle=90,fangle=-90](0)(0,0){}
  \drawloop[loopangle=180](0){$\begin{array}{l}
      a,\gfresh{1} \\ a,\rreg{1} \\
      a,\gfresh{2} \\ a,\rreg{2}
    \end{array}$}
\end{gpicture}

\begin{exa}\label{ex:2bounded}
  Consider the session automaton from
  \figurename~\ref{fig:multiples2}(b). It recognizes the
  set of all $2$-bounded data words over $\Sigma = \{a\}$.
\end{exa}

\subsection{Deterministic Session Automata}

Session automata come with two natural notions of determinism.  We
call $\A=(\States,\Reg,\init,\fStates,\Trans)$ \emph{symbolically
  deterministic} if $|\{s'\in\States\mid (s,(a,\param),s')\in\Trans\}|
\leq 1$ for all $s \in \States$, $a \in \Sigma$, and
$\param\in\gReg\cup\rReg$. Then, $\Trans$ can be seen as a partial
function $\States \times (\Sigma \times (\gReg\cup\rReg)) \pto S$.

We call $\A$ \emph{data deterministic} if it is symbolically
deterministic and, for all $s \in \States$, $a \in \Sigma$, and
$r_1,r_2 \in \Reg$ with $r_1 \neq r_2$, we have that
$(s,(a,\gfresh{r_1})) \in \domain \Trans$ implies
$(s,(a,\gfresh{r_2})) \notin \domain\Trans$. Intuitively, given a data
word as input, the automaton is data deterministic if, in each state,
given a pair letter/data value, there is at most one fireable
transition. 

Notice that session automata, even when symbolically or data
deterministic, may not necessarily be ``complete'', in the sense that
it is possible that a run over a data word falls into a deadlock
situation: this is the case when the session automaton forced a data
value to be removed from the set of registers, though it will be seen in
the future.

While ``data deterministic'' implies ``symbolically deterministic'' by
definition, the converse is not true. E.g., the session automaton
$\A_2$ from \figurename~\ref{fig:multiples3}(b) and
the one of \figurename~\ref{fig:multiples2}(b) are
symbolically deterministic but not data deterministic. However, the
session automaton
obtained from $\A_2$ by removing, e.g., the transition from $s_0$ to
$s_2$ (coupled with the transition from $s_0$ to $s_1$, it causes
non-determinism when reading a fresh data value at a request), is data
deterministic (and is indeed equivalent to $\A_2$, in the sense that
it recognizes the same language $L(\A_2)$).

\begin{exa}\label{ex:NFk}
  We explain how to construct a symbolically deterministic session
  automaton $\A$, with $k \ge 1$ registers, such that $\symbL(\A)=\NF_{k}$. Its state space is
  $\States=\{0,\ldots,k\}\times \powerset{\set{k}}$, consisting of
  (i) the greatest register already initialized (indeed we will only
  use a register $r$ if every register $r'<r$ has already been used),
  (ii) a subset $\Prom$ of registers that we promise to reuse again
  before resetting their value.  The initial state of $\A$ is
  $(0,\emptyset)$, whereas the set of accepting states is
  $(\{0,\ldots,k\})\times \{\emptyset\}$.  We now describe the set
  of transitions. For every $a\in \Sigma$, $i\in\{0,\ldots,k\}$,
  $\Prom\subseteq \set{k}$, and $r\in\set{k}$:
  \begin{align*} \Trans\big((i,\Prom),(a,\rreg{r})\big) &=
    \begin{cases} (i,\Prom\setminus\{r\}) &\text{if } r\leq i\\
      \text{not defined} &\text{otherwise}
    \end{cases}\\ \Trans\big((i,\Prom),(a,\gfresh{r})\big) &=
    \begin{cases} (\max(i,r),\Prom\cup\set{r-1}) &\text{if } r-1\leq i
      \land r\notin \Prom\\ \text{not defined} &\text{otherwise}
    \end{cases}
  \end{align*}
  \figurename~\ref{fig:NF2} depicts the session automaton for $\NF_{2}$ (omitting $\Sigma$).
  \begin{gpicture}[name=NF2,ignore]
    \unitlength=4 %
    \gasset{AHangle=30,ilength=3.5,flength=3.5,Nw=7,Nh=5,Nmr=5}
    \node[Nmarks=if,fangle=-90](0)(0,0){$0,\emptyset$}%
    \node[Nmarks=f,fangle=-90](1)(15,0){$1,\emptyset$}%
    \node[Nw=8,Nh=5,Nmr=5](3)(30,0){$2,\{1\}$}%
    \node[Nmarks=f,fangle=-90](5)(45,0){$2,\emptyset$}%
   
    \gasset{Nw=5,Nh=5,Nmr=5,loopdiam=4}
   
    \drawedge(0,1){$\gfresh{1}$}%
    \drawloop[loopangle=90](1){$\begin{array}{c} \gfresh{1} \\
        \rreg{1}
      \end{array}$} %
    \drawedge(1,3){$\gfresh{2}$}%
    \drawloop[loopangle=90](3){$\begin{array}{c} \gfresh{2} \\ \rreg{2}
      \end{array}$}%
    \drawedge[curvedepth=2](3,5){$\rreg{1}$}%
    \drawedge[curvedepth=2](5,3){$\gfresh{2}$}%
    \drawloop[loopangle=90](5){$\begin{array}{c} \gfresh{1} \\
        \rreg{1} \\ \rreg{2} \end{array}$}%
  \end{gpicture}
  \begin{figure}
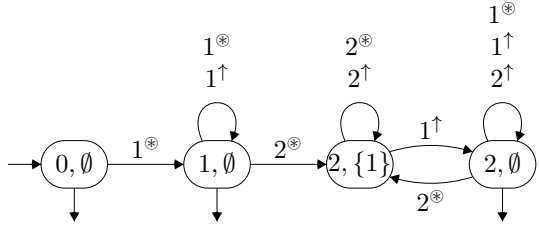
 
    \centering 
    \scalebox{.9}{\gusepicture{NF2}}
    \caption{A session automaton recognizing $\NF_{2}$\label{fig:NF2}}
  \end{figure}
\end{exa}

By determinizing a finite-state automaton recognizing the symbolic
language, it is easy to show that every language recognized by a
session automaton is also recognized by a symbolically deterministic
session automaton: we shall study this question in more detail in the
next section. The next theorem shows that this is not true for data
deterministic session automata.

\begin{thm}\label{thm:expressiveness}
  Session automata are strictly more expressive than data
  deterministic session automata.
\end{thm}
\proof
  We show that the data language $L = \Bounded{2}$ cannot
  be recognized by a data deterministic session automaton.  Indeed,
  suppose that such an automaton exists, with $k$ registers.  Then,
  consider the word $w=(a,1)(a,2)(a,3)\cdots (a,k+1) \in L$, where
  every data value is fresh.  By data determinism, there is a unique
  run accepting $w$.  Along this run, let $i<j$ be two positions such
  that their two fresh data values have been stored in the same
  register $r$ (such a pair must exist since the automaton has only
  $k$ registers). Without loss of generality, we can consider the
  greatest position $j$ verifying this condition, and then the
  greatest position $i$ associated with $j$. This means that register
  $r$ is used for the last time when reading $j$, and has not been
  used in-between positions $i$ and $j$.  Now, the word
  $(a,1)(a,2)(a,3)\cdots (a,k+1)(a,i) \in L$ must be recognized by the
  automaton, but cannot since data value $i$ appearing on the last
  position is not fresh anymore, and yet not stored in one of the
  registers (since register $r$ was reused at $j$). \qed

\subsection{Canonical Session Automata}

We now present the main result of this section showing that every
session automaton $\A$ is equivalent to a \emph{canonical} session
automaton $\can{\A}$, whose symbolic language $\symbL (\can\A)$
contains only symbolic normal forms.

\begin{thm}\label{thm:saregular}
  Let $\A = (\States,\Reg,\init,\fStates,\Trans)$ be a session
  automaton with $\Reg = \{1,\ldots,k\}$. Then, $L(\A)$ is
  $k$-bounded. Moreover, $\snf(L(\A))$ is a regular language over the
  finite alphabet $\Sigma \times \kNat{k}$. A corresponding
  automaton $\tilde\A$ can be effectively computed. Its number of
  states is at most exponential in $k$ and linear in $|\States|$.
\end{thm}
\proof
  First, if $\A$ is a session automaton using $k$ registers, the
  language $L(\A)$ is $k$-bounded since $\symbL(\A)\subseteq
  (\Sigma\times \kNat k)^\ast$, which implies that
  $L(\A)=\concretization(\symbL(\A))\subseteq
  \concretization((\Sigma\times \kNat k)^\ast)=\Bounded k$.
  
  Example~\ref{ex:NFk}, constructing a symbolically deterministic
  session automaton for $\NF_k=\snf(\concretization((\Sigma\times
  \kNat k)^\ast))$, shows that regularity of the symbolic language
  $(\Sigma\times \kNat{k})^{*}$ is preserved under the application of
  $\snf(\concretization(\cdot))$. We now prove that this is the case
  for every regular language over $\Sigma \times \Gamma_{k}$. In
  particular, for the symbolic \emph{regular} language $\symbL(\A)$,
  this will show that $\snf(L(\A))$, which is equal to
  $\snf(\concretization(\symbL(\A)))$, is regular.

  Let $L \subseteq (\Sigma \times \Gamma_k)^\ast$ be regular.
  Consider first the language
  \[\tilde L = \{u\in\WF \cap (\Sigma \times
  \Gamma_k)^\ast \mid \text{there is } u'\in L \text{ such that }
  \concretization(u)=\concretization(u')\}\] %
  i.e., the set of well-formed symbolic words having the same
  concretizations as some word from $L$.  We show that
  $\snf(\concretization(L)) = \NF_{k}\cap \tilde L$.  Indeed, if
  $u\in\snf(\concretization(L))$, then there are $u'\in L$ and
  $w\in\concretization(u')$ such that $u=\snf(w)$. Since
  $u'\in(\Sigma\times\kNat k)^\ast$, we have
  $u\in\snf(\concretization((\Sigma\times \kNat k)^\ast))=\NF_k$.
  Moreover, we have $[w]_{\approx} = \concretization(u')$ and
  $w\in\concretization(\snf(w))=\concretization(u)$ implying also
  $[w]_{\approx}=\concretization(u)$.  Finally, we obtain
  $\concretization(u)=\concretization(u')$, so that $u\in \tilde L$.
  Reciprocally, if $u\in\NF_{k} \mathrel{\cap} \tilde L$, then there
  is $u'\in L$ such that
  $\concretization(u)=\concretization(u')$. Hence, starting from a
  word $w$ in $\concretization(u)$ (which is non empty since $u$ is
  well-formed), we have $u=\snf(w)$ (by uniqueness of the symbolic
  normal form) and $w\in\concretization(u')\subseteq
  \concretization(L)$, so that $u\in\snf(\concretization(L))$.

  We know from Example~\ref{ex:NFk} that $\NF_{k}$ is regular. We now
  show that $\tilde L$ is regular: knowing that
  $\snf(\concretization(L))=\NF_k\cap \tilde L$, this will permit to
  conclude that $\snf(\concretization(L))$ is regular. To do so, let
  $\A=(\States,\Reg,\init,\fStates,\Trans)$ be a session automaton
  with $\Reg = \{1,\ldots,k\}$ such that $\symbL(\A)=L$.  We construct
  a session automaton $\tilde\A=(\States\times \Injection(k),\Reg,
  (s_{0},\emptyInj), \fStates\times \Injection(k), \tilde\Trans)$
  recognizing the symbolic language $\tilde L$.  Hereby,
  $\Injection(k)$ is the set of partial injective mappings from
  $\set{k}$ to $\set{k}$, and $\emptyInj \in \Injection(k)$ denotes
  the mapping with empty domain.  These partial mappings are used to
  remember the correspondence between old registers and new ones, so
  they may be understood as a set of constraints. For example, the
  mapping $(2\mapsto 1, 1\mapsto 3)$ stands for ``old register 2
  henceforth refers to 1, and old register 1 henceforth refers to
  3''. Each subset of these constraints forms always a \emph{valid}
  partial injective mapping. In the following, such a subset is called
  a sub-mapping. For example, $\sigma = (1\mapsto 3)$ is a sub-mapping
  of the previous one; it can then be extended with the new constraint
  $2\mapsto 2$, which we denote $\sigma[2\mapsto 2]$.  We describe now
  the transition relation of $\tilde\A$:
  \begin{align*}
    \tilde\Trans =
    &\phantom{{}\cup{}}\big\{\big((s_1,\sigma),(a,\rreg{\sigma(r)}),
    (s_2,\sigma)\big)
    \mid \big(s_1,(a,\rreg{r}),s_2\big)\in\Trans\big\}\\
    & {}\cup
    \big\{\big((s_1,\sigma_1),(a,\gfresh{r_2}),(s_2,\sigma_2)\big)
    \mid \big(s_1,(a,\gfresh{r_1}),s_2\big)\in\Trans
    \land \sigma_2 = \sigma[r_1\mapsto r_2]{}\\
    & \hspace{5em}\text{ with } \sigma \text{ maximal sub-mapping of }
    \sigma_1~ \text{s.t.\ } \sigma[r_1\mapsto r_2] \text{ injective}\big\}
  \end{align*}
  We simulate $\rreg{r}$-transitions simply using the current mapping
  $\sigma$. For $\gfresh{r}$-transitions, we update $\sigma$,
  recording the new permutation of the registers: the maximal
  sub-mapping $\sigma$ of $\sigma_1$ is either $\sigma_1$ itself or
  $\sigma_1$ where exactly one constraint $r_1\mapsto r_3$ is removed
  to free $r_1$. One can indeed show that $\symbL(\tilde\A) = \tilde
  L$. Inclusion $\symbL(\tilde\A) \subseteq \tilde L$ is easy to show
  since an accepting run in $\tilde\A$ can be mapped to an accepting
  run in $\A$ using the partial injective mappings maintained in the
  states of $\tilde\A$. For the other inclusion, it suffices to prove
  that for every symbolic word $u\in L$ and well-formed word $u'$ such
  that $\concretization(u')=\concretization(u)$, we have
  $u'\in\symbL(\tilde\A)$. By definition of $\concretization$, we know
  that projections of $u$ and $u'$ over the finite alphabet $\Sigma$
  are the same, and that ${\sim_u}={\sim_{u'}}$: the latter permits to
  reconstruct by induction a unique sequence of partial injective
  mappings linking the registers used in $u$ and in $u'$. An accepting
  run of $\A$ on $u$ can therefore be mapped to an accepting run of
  $\tilde\A$ on $u'$. 

  Building the product of the automaton recognizing $\NF_{k}$ and the
  automaton $\tilde\A$, we obtain a session automaton using $k$
  registers recognizing $\snf(\concretization(L))$. Its number of
  states is bounded above by $O(|Q|\times k!\times (k+1)\times 2^{k})$
  (as the number of partial injective mappings in $\Injection(k)$ is
  bounded above by $O(k!)$). \qed

From the automaton $\tilde\A$ built in the proof of the previous
theorem, we can consider the (unique up to isomorphism) minimal
deterministic finite-state automaton $\can\A$ (i.e., symbolically
deterministic session automaton) equivalent to it: this automaton will
be called the \emph{canonical session automaton}. In case $\A$ is data
deterministic, we can verify that $\tilde\A$ is symbolically
deterministic, and hence the minimal automaton $\can\A$ has at
most $O(|Q|\times k!\times (k+1)\times 2^{k})$ states. Otherwise, a
determinization phase has to be done resulting in a canonical session
automaton with at most $2^{O(|Q|\times k!\times (k+1)\times 2^{k})}$
states.

\begin{exa} Examples of $\A$ and $\tilde\A$, as defined in the
  previous proof, are given in \figurename~\ref{fig:exampleCanonical}.
  The figure also depicts the canonical automaton $\can\A$ associated
  with $\A$, obtained by determinizing and minimizing the product of
  both $\tilde\A$ and the symbolically deterministic automaton recognizing
  $\NF_{2}$ (as given in \figurename~\ref{fig:NF2}). Note that
  $\can\A$ is symbolically deterministic and minimal.
  \begin{gpicture}[name=exampleCanonical1,ignore]
    \unitlength=4 \gasset{Nw=5,Nh=5,Nmr=5,loopdiam=4}
    \gasset{AHangle=30,ilength=3.5,flength=3.5}
    \node[Nmarks=if,iangle=90,fangle=-90](0)(0,0){}
    \drawloop[loopangle=0](0){$\begin{array}{l}
        a,\gfresh{1} \\ b,\rreg{1} \\
        a,\gfresh{2} \\ b,\rreg{2}
      \end{array}$}
  \end{gpicture}
  \begin{gpicture}[name=exampleCanonical2,ignore]
    \unitlength=5 \gasset{Nw=5,Nh=5,Nmr=5,loopdiam=3}
    \gasset{AHangle=30,ATangle=30,ilength=3.5,flength=3.5}
    \node[Nmarks=if,iangle=90,fangle=30](0)(0,0){{\scriptsize$\emptyInj$}}
    \gasset{Nw=7}
    \node[Nmarks=f,fangle=-120](1)(8,-15){{\scriptsize$1\mapsto 1$}}
    \node[Nmarks=f,fangle=-60](2)(-8,-15){{\scriptsize$2\mapsto 1$}}
    \node[Nmarks=f,fangle=0](1')(25,-15){{\scriptsize$1\mapsto 2$}}
    \node[Nmarks=f,fangle=180](2')(-25,-15){{\scriptsize$2\mapsto 2$}}
    \gasset{Nw=9}
    \node[Nmarks=f,fangle=180](3)(13,-30){{\scriptsize$\begin{array}{c}1\mapsto
          1\\2\mapsto 2\end{array}$}}
    \node[Nmarks=f](4)(-13,-30){{\scriptsize$\begin{array}{c}1\mapsto
          2\\2\mapsto 1\end{array}$}}
    \drawedge[ELside=r,ELpos=40,ELdist=0](0,1){$a,\gfresh{1}$}
    \drawedge[ELpos=60,ELdist=0](0,2){$a,\gfresh{1}$}
    \drawedge[curvedepth=5,ELpos=30](0,1'){$a,\gfresh{2}$}
    \drawedge[curvedepth=-5,ELside=r,ELpos=30](0,2'){$a,\gfresh{2}$}
    \drawloop[loopangle=70,ELdist=0](1){$\begin{array}{l} a,\gfresh{1}\\
        b,\rreg{1}
      \end{array}$}
    \drawloop[loopangle=120,ELdist=0](2){$\begin{array}{l}
        a,\gfresh{1} \\ b,\rreg{1}
      \end{array}$}
    \drawloop[loopangle=63,ELdist=0](1'){$\begin{array}{l}
        a,\gfresh{2}\\ b,\rreg{2}
      \end{array}$}
    \drawloop[loopangle=120,ELdist=0](2'){$\begin{array}{l}
        a,\gfresh{2} \\ b,\rreg{2}
      \end{array}$}
    \drawedge[ATnb=1](1,2){$a,\gfresh{1}$}
    \drawedge[curvedepth=2](1,1'){$a,\gfresh{2}$}
    \drawedge[curvedepth=2](1',1){$a,\gfresh{1}$}
    \drawedge[curvedepth=2](2,2'){$a,\gfresh{2}$}
    \drawedge[curvedepth=2](2',2){$a,\gfresh{1}$}
    \drawbpedge[ATnb=1,ELside=r](1',100,30,2',80,30){$a,\gfresh{2}$}
    \drawedge[curvedepth=0](1,3){$a,\gfresh{2}$}
    \drawedge[curvedepth=0,ELside=r](2,4){$a,\gfresh{2}$}
    \drawbpedge(1',-50,30,4,-50,20){$a,\gfresh{1}$}
    \drawbpedge[ELside=r](2',-130,30,3,-130,20){$a,\gfresh{1}$}
    \drawloop[loopangle=10,ELdist=0](3){$\begin{array}{l} a,\gfresh{1} \\
        a,\gfresh{2} \\ b,\rreg{1} \\b,\rreg{2}
      \end{array}$}
    \drawloop[loopangle=170,ELdist=0](4){$\begin{array}{l}
        a,\gfresh{1} \\ a,\gfresh{2} \\ b,\rreg{1} \\b,\rreg{2}
      \end{array}$}
    \drawedge[ELpos=30,ELdist=0.1](3,2){$a,\gfresh{1}$}
    \drawedge[ELpos=30,ELside=r,ELdist=0.1](4,1){$a,\gfresh{1}$}
    \drawedge[ELpos=70,ELside=r,ELdist=0.1](3,1'){$a,\gfresh{2}$}
    \drawedge[ELpos=70,ELdist=0.1](4,2'){$a,\gfresh{2}$}
  \end{gpicture}
  \begin{gpicture}[name=exampleCanonical3,ignore]
    \unitlength=3 \gasset{Nw=5,Nh=5,Nmr=5,loopdiam=4}
    \gasset{AHangle=30,ilength=3.5,flength=3.5}
    \node[Nmarks=if,iangle=90,fangle=180](0)(0,0){}
    \node[Nmarks=f,fangle=180](1)(0,-15){} \node(3)(0,-30){}
    \node[Nmarks=f,fangle=180](5)(0,-45){}
    
    \drawedge(0,1){$a,\gfresh{1}$}
    \drawloop[loopangle=0](1){$\begin{array}{l} a,\gfresh{1} \\ b,\rreg{1}
      \end{array}$}
    \drawedge(1,3){$a,\gfresh{2}$}
    \drawloop[loopangle=0](3){$\begin{array}{l}
        a,\gfresh{2} \\ b,\rreg{2}
      \end{array}$}
    \drawedge[curvedepth=2](3,5){$b,\rreg{1}$}
    \drawedge[curvedepth=2](5,3){$a,\gfresh{2}$}
    \drawloop[loopangle=0](5){$\begin{array}{l}
        a,\gfresh{1} \\ b,\rreg{1} \\ b,\rreg{2}
      \end{array}$}
  \end{gpicture}
  \begin{figure}[t]
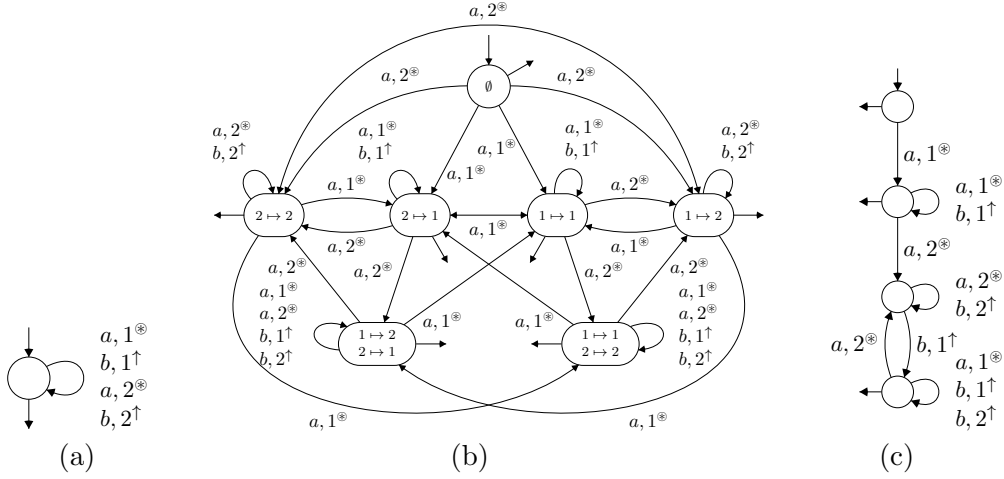

    \centering
    \begin{tabular}{ccc}
    \scalebox{.8}{\gusepicture{exampleCanonical1}} &\hspace{1em}
    \scalebox{.65}{\gusepicture{exampleCanonical2}}&\hspace{1em}
    \scalebox{.8}{\gusepicture{exampleCanonical3}}\\
    (a)&(b)&(c)
    \end{tabular}
    \caption{(a) A session automaton $\A$, (b) its automaton
      $\tilde\A$, (c) its canonical automaton
      $\can\A$\label{fig:exampleCanonical}}
  \end{figure}
\end{exa}

\subsection{Closure Properties}

Using Theorem~\ref{thm:saregular}, we obtain some language theoretical
closure properties of session automata, which they inherit from
classical regular languages. These results demonstrate a certain
robustness as required in verification tasks such as compositional
verification \cite{CobleighGP03} and infinite-state regular model
checking \cite{HV-infinity04}.

\begin{thm}
We have the following closure properties:
\begin{itemize}\label{thm:closure}
\item Session automata are closed under union and intersection.
\item Session automata are closed under resource-sensitive complementation:
  Given a session automaton $\A$ with $k$ registers, there is a
  session automaton $\A'$ with $k$ registers such that $L(\A')
  = \Bounded{k} \setminus L(\A)$.
\end{itemize}
\end{thm}
\proof
  Let $\A$ be a session automaton using $k$ registers, and $\B$ a
  session automaton using $k'$ registers. Using a classical product
  construction for $\can \A$ and $\can \cB$, we obtain a session
  automaton using $\min(k,k')$ registers recognizing the data language
  $L(\A)\cap L(\B)$.  The language $L(\A)\cup L(\B)$
  is recognized by the session automaton, using $\max(k,k')$
  registers, that we obtain as the ``disjoint union'' of $\A$ and
  $\cB$, branching on the first transition in one of these two
  automata.

  Finally, let us consider a symbolically deterministic session
  automaton $\A$ using $k$ registers. Without loss of generality, by
  adding a sink state, we can suppose that $\A$ is complete. Then,
  every well-formed symbolic word over $\Sigma\times \kNat{k}$ has
  exactly one run in $\A$. The automaton $\A'$ constructed from $\A$
  by taking as accepting states the non-accepting states of $\A$
  verifies that $\symbL(\A') = (\Sigma\times
  \kNat{k})^{*}\setminus\symbL(\A)$ so that $L(\A')
  =\concretization((\Sigma\times \kNat{k})^{*})\setminus
  L(\A)$. Notice that $\A'$ is symbolically deterministic, but not
  necessarily data deterministic (even if $\A$ is), because of the
  completion step.\qed

\begin{thm}\label{thm:inclusion}
  The inclusion problem for session automata is decidable.
\end{thm}
\proof Considering two session automata $\A$ and $\B$, we can decide
inclusion $L(\A)\subseteq L(\B)$ by considering the canonical automata
$\can\A$ and $\can\B$. Indeed, $L(\A)\subseteq L(\B)
\Longleftrightarrow \snf(L(\A))\subseteq \snf(L(\B))
\Longleftrightarrow \symbL(\can\A)\subseteq \symbL(\can\B)$.
Thus, it is sufficient to check inclusion for $\can\A$ and
$\can\B$. \qed

In case $\B$ is data deterministic, $\can\B$ has a size polynomial in
the number of states of $\B$, but exponential in the number of
registers. Testing the inclusion $\symbL(\can\A)\subseteq
\symbL(\can\B)$ may be done by first complementing $\symbL(\can\B)$
(which does not add states since $\can\B$ is symbolically
deterministic) and then testing the emptiness of its intersection with
$\symbL(\can\A)$. In the overall, this implies a complexity of the
inclusion check that is polynomial in the number of states of $\A$ and
$\B$, but exponential in the number of registers used by $\B$. In case
$\B$ is not data deterministic, a determinization phase may add an
exponent in the size and the number of registers of $\B$.

As a corollary, we obtain that the emptiness problem and the
universality problem under $k$-boundedness (i.e., knowing whether the
language of a session automaton with $k$ registers is the whole set of
$k$-bounded data words) are decidable for session automata. It is not
surprising for the emptiness problem, since it already holds for
fresh-register automata. Notice that the problem is shown
co-NP-complete for register automata in \cite{SakIke00},
and we can show that the
emptiness problem is co-NP-complete, too, for session automata. First,
co-NP-hardness can be shown by a reduction to the 3-SAT problem, in a
very similar way as in \cite{SakIke00}. Then, the co-NP upper bound
comes from the symbolic view. Indeed, for a session automaton $\A$
with $k$ registers, $L(\A)=\emptyset$ if and only if $\symbL(\A) \cap
\WF_k=\emptyset$ (where $\WF_k$ denotes the set of well-formed
symbolic words over alphabet $\Sigma\times \Gamma_k$). We may not
construct a finite automaton recognizing $\symbL(\A) \cap \WF_k$ (that
has a size exponential in $k$), but instead non-deterministically
search for a witness of non-emptiness of $\symbL(\A) \cap \WF_k$,
i.e., a well-formed word $u$ such that $u\in\symbL(\A)$. Notice that
the membership test of $u$ in the finite-state automaton $\A$ can be
performed in polynomial time, hence, to conclude, we must simply show
the existence of a well-formed witness $u$ of polynomial size. As for
register automata in \cite{SakIke00}, this relies on the fact that,
even though the total number of configurations of $\A$ is exponential
in $k$ (due to the set $U$ of initialized registers), along a run of
$\A$, only a polynomial (in $\A$ and $k$) number of configurations can
be visited, since the set $U$ will take at most $k+1$ values during
the computation (the initialization of registers is done in a certain
order, and no register can be emptied at any point). Hence, by
disallowing the visit of two occurrences of the same configuration,
the existence of a well-formed witness implies the existence of a
well-formed witness of polynomial size.

While fresh-register automata are not complementable for the set of
all data words, they are complementable for $k$-bounded data words,
using the previous theorem. The reason is that, given a fresh-register
automaton $\A$, one can construct a session automaton $\B$ such that
$L(\B) = L(\A) \cap \Bounded{k}$.

\section{Logical Characterizations}

\label{sec:logic}

In this section, we provide logical characterizations of session automata.

\subsection{MSO Logic over Data Words}

We consider the standard \emph{data monadic second-order logic} (dMSO), which is an extension of
classical MSO logic by the binary predicate $x \sim y$ to compare data
values.  

We fix infinite supplies of first-order variables $x,y, \ldots$, which
are interpreted as word positions, and second-order variables $X,Y,
\ldots$, which are taken as sets of positions.
We let dMSO be the set of formulae $\phi$ defined by the following
grammar:
\[\phi ~~::=~~ \mylabel(x)=a \mid x=y \mid y=x+1\mid x\sim
y\mid x\in X\mid \neg \phi \mid \phi\lor\phi \mid \exists x \; \phi\mid \exists
X\; \phi\] with $x,y$ first-order variables, $X$ a second-order variable
and $a\in\Sigma$. The semantics of formulae in dMSO is given in
Table~\ref{tab:semantics}: we define $w,\sigma\models \phi$ (to be
read as ``$w$ satisfies $\phi$ when free variables of $\phi$ are
interpreted as prescribed in $\sigma$'') by induction over $\phi$,
where $w=(a_1,d_1)\cdots (a_n,d_n) \in (\Sigma \times D)^\ast$ is a
data word and $\sigma$ is a valuation of (at least the) free variables in $\phi$,
i.e., such that for every first-order free variable $x$, we have
$\sigma(x)\in\set{n}$ and for every second-order free variable $X$, we
have $\sigma(X)\subseteq \set{n}$. For a first-order variable $x$ and
a position $i\in\set{n}$, we let $\sigma[x\mapsto i]$ be the valuation
$\tau$ such that $\tau(x)=i$ and $\tau(\alpha)=\sigma(\alpha)$ for
every variable $\alpha$ different from $x$. A similar definition holds for
second-order variables.

\begin{table}[tbp]
  \caption{Semantics of formulae in dMSO}
  \centering
  $\begin{array}{r @{\quad} l}
    w,[x\mapsto i,y\mapsto j]\models x=y & \text{if } i=j \\
    w,[x\mapsto i]\models \mylabel(x)=a & \text{if } a_i=a\\
    w,[x\mapsto i,y\mapsto j]\models y=x+1 & \text{if } j=i+1\\
    w,[x\mapsto i,X\mapsto I]\models x\in X & \text{if } i\in I\\
    w,[x\mapsto i,y\mapsto j]\models x\sim y & \text{if } d_i=d_j\\
    w,\sigma \models \neg \phi & \text{if } w,\sigma \not\models \phi
    \\
    w,\sigma \models \phi_1\lor \phi_2 & \text{if } w,\sigma\models
    \phi_1 \text{ or } w,\sigma\models \phi_2\\
    w,\sigma \models \exists x\; \phi & \text{if there exists }
    i\in\set{n} \text{ such that } w,\sigma[x \mapsto i]\models \phi\\
    w,\sigma \models \exists X\; \phi & \text{if there exists }
    I\subseteq\set{n} \text{ such that } w,\sigma[X \mapsto I]\models \phi
  \end{array}$
  \label{tab:semantics}
\end{table}

In addition, we use abbreviations such as $\ttrue$, $x \le y$,
$\forall x\; \phi$, $\phi \wedge \psi$, $\phi \rightarrow \psi$,
etc. A sentence is a formula without free variables. For a dMSO
sentence $\phi$, we set $L(\phi) \df \{w \in (\Sigma \times D)^\ast
\mid w \models \phi\}$. Note that $L(\phi)$ is a data language.

As usual, to deal with free variables, it is possible to extend the
alphabet $\Sigma$ as follows. If $V$ is the set of variables that
occur in $\phi$, we have to consider data words over $\hat{\Sigma} =
\Sigma \times \{0,1\}^V$ and $\Data$.  Intuitively, these data words
include the interpretation of the free variables. If a data word
carries, at position $i$, the letter $(a,\bar{b},d) \in \hat{\Sigma}
\times \Data$ with $\bar{b}[x] = 1$ (where $\bar{b}[x]$ refers to the
$x$-component of $\bar{b}$), then $x$ is interpreted as position
$i$. If $\bar{b}[X] = 1$, then $X$ is interpreted as a set
\emph{containing} $i$.  Whenever we refer to a word over the extended
alphabet $\hat{\Sigma}$, we will silently assume that the
interpretation of a first-order variable $x$ is uniquely determined,
i.e., there is exactly one position $i$ where $\bar{b}[x] = 1$. This
is justified, since the set of those ``well-shaped'' words is
(symbolically) regular. This way we can transform any well-shaped word
$\hat w\in (\Sigma \times \{0,1\}^V\times \Data)^*$ into a pair
$(w,\sigma)$ where $w$ is a data word of $(\Sigma\times \Data)^*$ and
$\sigma$ is a valuation of variables in $V$, and vice versa.

Note that dMSO is a very expressive logic and goes beyond virtually
all automata models defined for data words
\cite{Neven2004,Segoufin06,BojanczykDMSS11,Colcombet2011}.  However,
if we restrict to bounded languages, we can show that dMSO
is no more expressive than session automata.

\begin{thm}\label{thm:dMSO}
  Let $L$ be a bounded data language. Then, the following statements are
  equivalent:
  \begin{itemize}
  \item There is a session automaton $\A$ such that $L(\A) = L$.
  \item There is a dMSO sentence $\phi$ such that $L(\phi) = L$.
  \end{itemize}
\end{thm}
\proof The construction of a dMSO formula of the form $\exists X_1
\cdots \exists X_m\; (\runf \wedge \forall x\forall y\;(x \sim y
\leftrightarrow \dataf))$, with $\runf$ and $\dataf$ classical MSO
formulae (not containing predicate $\sim$), from a session automaton
$\A$ was implicitly shown in \cite{BCGK-fossacs12} (with a different
goal, though). The idea is that the existential second-order variables
$X_1,\ldots,X_m$ are used to guess an assignment of transitions to
positions. In $\runf$, it is verified that the assignment corresponds
to a run of $\A$. Moreover, $\dataf$ checks if data equality
corresponds to the data flow as enforced by the transition labels from
$\Gamma_k$. The formula has a size polynomial in the size of the
automaton. In Section~\ref{sec:sMSO}, formulae of this shape will be studied
in more detail.

For the converse direction, we perform, as usual, an induction on the
structure of the formula $\phi$ of dMSO such that $L(\phi)$ is
$k$-bounded, for some $k \ge 1$. To cope with free variables, we use the
encoding of a pair $(w,\sigma)$ as presented before.

First, we have to deal with the base cases:
\begin{itemize}
\item Consider the formula $\mylabel(x) = a$. We
construct a session automaton $\A$ with $k$ registers such that
$\symbL(\A)$ consists of all ``well-shaped'' words $u \in
(\hat{\Sigma} \times \Gamma_k)^\ast$ containing a letter
$(a,\bar{b},\pi)$ with $\bar{b}[x] = 1$.
\item For $x = y$, the automaton has to verify that there is a letter
  $(a,\bar{b},\pi)$ such that $\bar{b}[x] = \bar{b}[y] = 1$, which can
  be done since this is a regular condition on word over alphabet
  $\hat\Sigma\times \Gamma_k$.
\item Formulae $y = x+1$ and $x \in X$ are treated similarly.
\item The most interesting base case is $x \sim y$. Let $L$ be the
  symbolic language containing exactly the symbolic words
  $u=(a_1,\bar{b}_1,\pi_1) \cdots (a_n,\bar{b}_n,\pi_n) \in
  (\hat{\Sigma} \times \Gamma_k)^\ast$ satisfying the following: there
  are two positions $i,j \in \{1,\ldots,n\}$ such that $i \sim_u j$,
  $\bar{b}_i[x] = 1$, and $\bar{b}_j[y] = 1$. Note that $L$ is indeed
  a regular language so that we can construct a corresponding session
  automaton $\A$ with $k$ registers such that $\symbL(\A)=L$.
\end{itemize}

\noindent In all of the above base cases, if $\phi$ is the atomic formula and
$\A$ the corresponding session automaton, the following holds: given a
data word $\hat w \in \Bounded{k}$ encoding free variables, we have
$\hat w \in L(\phi)$ iff $\hat w \in L(\A)$.

\smallskip
Let us come to the induction step. To deal with negation, we can
indeed exploit the fact that session automata are closed under
complementation when considering only $k$-bounded data words
(Theorem~\ref{thm:closure}). Suppose we have constructed a session
automaton $\A$ with $k$ registers such that $L(\A) = L(\phi) \cap
\Bounded{k}$.  According to Theorem~\ref{thm:closure}, there is a
session automaton $\A'$ with $k$ registers such that $L(\A') =
\Bounded{k} \setminus L(\A)$. From $L(\neg\phi)=(\Sigma\times D)^*
\setminus L(\phi)$, we deduce $L(\A') = L(\neg\phi) \cap
\Bounded{k}$. To deal with disjunction, we exploit closure of session
automata under union (again, Theorem~\ref{thm:closure}). Finally,
existential quantification corresponds, as usual, to projection. 

Because of the negations that require complementation (and hence
determinization of finite-state automata), the automaton associated
with a given dMSO formula has a size given as a tower of exponentials
of the size given by the number of nested negations in the formula. \qed

By Theorems~\ref{thm:inclusion} and \ref{thm:dMSO}, we obtain, as a
corollary, that model checking session automata against dMSO
specifications is decidable, though with non-elementary complexity
(while it is undecidable for register
automata). Note that this was already shown in \cite{BCGK-fossacs12}
for a more powerful model with pushdown stacks. 

\begin{thm}
  Given a session automaton $\A$ and a dMSO sentence $\phi$, one can
  decide whether $L(\A) \subseteq L(\phi)$.
\end{thm}

\subsection{Session MSO Logic}\label{sec:sMSO}

Next, we give an alternative logical characterization of session
automata.  We identify a fragment of dMSO, called \emph{session MSO
  logic}, such that every formula from that fragment can be translated
into a session automaton, without having to restrict the set of data
words in advance.  Note that register automata also enjoy a logical
characterization \cite{Colcombet2011}.  There, \emph{guards} are
employed to tame the power of the predicate $\sim$. Similarly, our
logic also uses a guard, though in a quite different way.

\begin{defi}
  A \emph{session MSO} (sMSO) formula is a dMSO sentence of the form
  \[\exists X_1 \cdots \exists X_m\; (\runf \wedge \forall
  x\forall y\; (x \sim y \leftrightarrow \dataf))\] such that $\runf$
  and $\dataf$ are classical MSO formulae (not containing the
  predicate $\sim$).
\end{defi}

\begin{exa}
  The formula $\phi_1 = \forall x\forall y\; (x \sim y
  \leftrightarrow x = y)$ is an sMSO formula.  Its semantics
  $L(\phi_1)$ is the set of data words in which every data value
  occurs at most once.  Moreover, $\phi_2 = \forall x\forall y\; (x
  \sim y \leftrightarrow \ttrue)$ is an sMSO formula, and
  $L(\phi_2)$ is the set of data words where all data values
  coincide.  As a last example, let $\phi_3 = \exists X\, \forall
  x\forall y\;(x \sim y \leftrightarrow (\neg\exists z \in X\; (x < z
  \le y \vee y < z \le x)))$.  Then, $L(\phi_3)$ is the set of
  $1$-bounded data words. Intuitively, the second-order variable $X$
  represents the set of positions where a fresh data value is
  introduced.
\end{exa}

\begin{thm}\label{thm:logic}
  For all data languages $L$, the following statements are
  equivalent:
  \begin{itemize}
  \item There is a session automaton $\A$ such that $L(\A) = L$.
  \item There is an sMSO sentence $\phi$ such that $L(\phi) = L$.
  \end{itemize}
\end{thm}

\proof The construction of an sMSO formula from a session automaton
$\A$ has already been sketched in the proof of Theorem~\ref{thm:dMSO}.

We turn to the converse direction and let $\phi = \exists X_1 \cdots
\exists X_m\;(\runf \mathrel{\wedge} \forall x\forall y\;(x \sim y
\leftrightarrow \dataf))$ be an sMSO sentence. By
Theorem~\ref{thm:dMSO}, it is sufficient to show that $L(\phi)$ is
bounded.

The free variables of $\dataf$ are among $x,y,X_1,\ldots,X_m$. As,
moreover, $\dataf$ is a ``classical'' MSO formula, which does not
contain $\sim$,
it defines, in the expected manner, a set $\symbL(\dataf)$ of words
over the finite alphabet $\Sigma \times \{0,1\}^{m+2}$.  Similarly to
the proof of Theorem~\ref{thm:dMSO}, the idea is to interpret a
position carrying letter $(a,1,b,b_1,\ldots,b_m)$ as $x$, and a
position labeled $(a,b,1,b_1,\ldots,b_m)$ as $y$, while membership in
$X_i$ is indicated by $b_i$. Words where $x$ and $y$ are not uniquely
determined, are discarded. We can represent such models as tuples
$(w,i,j,I_1,\ldots,I_m)$ where $w \in \Sigma^\ast$, $i$ denotes the
position of the $1$-entry in the unique first component, and $j$
denotes the position of the $1$-entry in the second component.
As $\symbL(\dataf) \subseteq (\Sigma \times \{0,1\}^{m+2})^\ast$ is
MSO definable (in the classical sense, without data), it is, by
B{\"u}chi's theorem, recognized by some minimal deterministic finite
automaton $\A_\dataf$. Suppose that $\A_\dataf$ has $\kphi \ge 1$
states.

We claim that the data language $L(\phi)$ is $\kphi$-bounded.
To show this, let $w = (a_1,d_1)$ $\cdots$ $(a_n,d_n) \in L(\phi)$.
There exists a tuple $\tupleI =(I_1,\ldots,I_m)$ of subsets of
$\set{n}$ such that, for all $i,j \in \set{n}$, \[d_i = d_j
~\Longleftrightarrow~ (a_1 \cdots a_n,i,j,\tupleI) \in
\symbL(\dataf)\,.\tag{$\ast$}\label{lab:deq}\]

Suppose, towards a contradiction, that $w$ is not
$\kphi$-bounded. Then, there are $k > \kphi$ and a position $i \in
\set{n}$ such that $i$ is contained in $k$ distinct sessions
$J_1,\ldots,J_{\newk}$. For $l \in \{1,\ldots,\newk\}$, let $i_l =
\min(J_l)$ and $j_l = \max(J_l)$, so that $J_l =
\{i_l,i_l+1,\ldots,j_l\}$. Note that the $i_l$ are pairwise distinct,
and so are the $j_l$. By~\eqref{lab:deq}, for every $l \in
\{1,\ldots,\newk\}$, we have $w_l = (a_1 \cdots a_n,i_l,j_l,\tupleI)
\in \symbL(\dataf)$. Thus, for every such word $w_l$, there is a
unique accepting run of $\A_\dataf$, say, being in state $q_l$ after
executing position $i$. As $\A_\dataf$ has only $\kphi$ states, there
are $l \neq l'$ such that $q_l = q_{l'}$. Thus, there is an accepting
run of $\A_\dataf$ either on a word where one of the first-order
components is not unique, which is a contradiction, or on $(a_1 \cdots
a_n,i_l,j_{l'},\tupleI)$. The latter contradicts (\ref{lab:deq}),
since $J_l$ and $J_{l'}$ are distinct sessions. \qed

\section{Learning Session Automata}
\label{sec:learning}

In this section, we introduce an active learning algorithm for session
automata. In the usual active learning setting (as introduced by
Angluin \cite{Angluin:regset}, see \cite{Hig10} for a general overview
of active learning techniques), a \emph{learner} interacts with a
so-called minimally adequate \emph{teacher} (MAT), an oracle which can
answer \emph{membership} and \emph{equivalence queries}.  In our case,
the learner is given the task to infer the data language $ L(\A)$
defined by a given session automaton $\A$.  We suppose here that the
teacher knows the session automaton or any other device accepting $
L(\A)$. In practice, this might not be the case --- $\A$ could be a
black box --- and equivalence queries could be (approximately)
answered, for example, by extensive testing.  The learner can ask if a
\emph{data} word is accepted by $\A$ or not.  Furthermore it can ask
equivalence queries which consist in giving an \emph{hypothesis}
session automaton to the teacher who either answers yes, if the
hypothesis is equivalent to $\A$ (i.e., both data languages are the
same), or gives a data word which is a counterexample, i.e., a data
word that is either accepted by the hypothesis automaton but should
not, or vice versa.

Given the data language $ L(\A)$ accepted by a session automaton $\A$
over $\Sigma$ and $D$, our algorithm will learn the canonical session
automaton $\can{\A}$, that uses $k$ registers, i.e., the minimal
symbolically deterministic automaton recognizing the language $L(\A)$
and the regular language $\symbL(\can{\A})$ over $\Sigma \times
\Gamma_{k}$.  Therefore one can consider that the learning target is
$\symbL(\can{\A})$ and use an arbitrary active learning algorithm for
regular languages. However, as the teacher answers only questions over
data words, queries have to be adapted. Since $\can{\A}$ only accepts
symbolic words which are in normal form, a membership query for a
given symbolic word $u$ not in normal form will be answered negatively
(without consulting the teacher); otherwise, the teacher will be given
one data word included in $\gamma(u)$ (all the answers on words of
$\gamma(u)$ are the same).  Likewise, before submitting an equivalence
query to the teacher, the learning algorithm checks if the current
hypothesis automaton accepts symbolic words not in normal
form\footnote{This can be checked in polynomial time over the trimmed
  hypothesis automaton with a fixed point computation labelling the
  states with the registers that should be used again before
  overwriting them.}. If yes, one of those is taken as a
counterexample, else an equivalence query is submitted to the
teacher. Since the number of registers needed to accept a data
language is a priori not known, the learning algorithm starts by
trying to learn a session automaton with $1$ register and increases
the number of registers as necessary.

Every active learning algorithm for regular languages may be adapted
to our setting.  Here we describe a variant of Rivest and Schapire's
algorithm \cite{RiSh:inference} which is itself a variant of Angluin's
L$^*$ algorithm \cite{Angluin:regset}.  An overview of learning
algorithms for deterministic finite state automata can be found, for
example, in \cite{BergR05}.

The algorithm is based on the notion of \emph{observation table} which
contains the information accumulated by the learner during the
learning process.  An observation table over a given alphabet $\Sigma
\times \Gamma_{k}$ is a triple $\mathcal O = (T,U,V)$ with $U,V$ two
sets of words over $\Sigma \times \Gamma_{k}$ such that $\varepsilon
\in U \cap V$ and $T$ is a mapping $(U \mathrel{\cup} U\cdot (\Sigma
\times \Gamma_{k})) \times V \rightarrow \{+,-\}$.  A table is
partitioned into an upper part $U$ and a lower part $U\cdot (\Sigma
\times \Gamma_{k})$.  We define for each $u \in U \mathrel{\cup}
U\cdot (\Sigma \times \Gamma_{k})$ a mapping $row(u)\colon V
\rightarrow \{+,-\}$ where $row(u)(v) = T(u,v)$.  An observation table
must satisfy the following property: for all $u,u' \in U$ such that $u
\not= u'$ we have $row(u) \not= row(u')$, i.e., there exists $v \in V$
such that $T(u,v) \not= T(u',v)$. This means that the rows of the
upper part of the table are pairwise distinct.  A table is
\emph{closed} if for all $u'\in U\cdot (\Sigma \times \Gamma_{k})$
there exists $u \in U$ such that $row(u) = row(u')$.  From a closed
table we can construct a symbolically deterministic session automaton
whose states correspond to the rows of the upper part of the table:

\begin{defi}
  \label{def:autfromtable}
  For a closed table $\mathcal O=(T,U,V)$ over a finite alphabet
  $\Sigma \times \Gamma_{k}$, we define a symbolically deterministic
  session automaton $A_{\mathcal O} =
  (\States,\Reg,\init,\fStates,\Trans)$ over $\Sigma \times
  \Gamma_{k}$ by $\States=U$, $\Reg = \{1,\ldots,k\}$, $\init=
  \varepsilon$, $\fStates=\{u \in \States \mid T(u,\varepsilon)=+\}$,
  and for all $u \in \States$ and $(a,\param) \in \Sigma \times
  \Gamma_{k}$, $\Trans(u,(a,\param)) = u'$ if $row(u(a,\param))=
  row(u')$. This is well defined as the table is closed.
\end{defi}

\setlength{\algomargin}{.2em} 
\begin{algorithm2e}[bt]\footnotesize
  initialize $k:=1$ and 
  $\mathcal O := (T,U,V)$ by $U=V=\{\varepsilon\}$
  and $T(u,\varepsilon)$ for all $u \in U \cup U\cdot
  (\Sigma \times \Gamma_{k})$ with membership queries\;

  \Repeat{equivalence test succeeds}{
    \While{$\mathcal O$ is not closed}{
      find  $u\in U$ and $(a,\param) \in \Sigma \times \Gamma_{k}$
      such that for all $u'\in U:~row(u(a,\param)) \neq row(u')$\;
      extend table to $\mathcal O := (T',U \cup \{u(a,\param)\}, V)$ by  membership queries\;
    }
    from $\mathcal O$ construct the hypothesized automaton $\A_{\mathcal O}$\tcp*[r]{cf.\
    Definition~\ref{def:autfromtable}}
    \eIf{$\A_{\mathcal O}$ accepts symbolic words not in normal form}
    {
      let $z$ be one of those\;
    }
    {
      \eIf{$ L(\A)= L(\A_{\mathcal O})$}
      {
        equivalence test succeeds\;
      }
      {
        get counterexample $w \in ( L(\A) \setminus  L(\A_{\mathcal O}))
        \cup ( L(\A_{\mathcal O}) \setminus  L(\A))$\; 
        set $z := \snf(w)$\; 
        find minimal $k'$ such that $z \in (\Sigma \times \Gamma_{k'})^\ast$\;
        \If{$k'>k$}
        {
          set $k := k'$\;
          extend table to $\mathcal O := (T',U,V)$ over $\Sigma \times \Gamma_{k}$ by
          membership queries\;
        }
      }
    }
    \If(\tcp*[f]{is true if $k' \leq k$}){$\mathcal O$ is closed}
    {
      find a break-point for $z$ where $v$ is the distinguishing word\;
      extend table to $\mathcal O :=(T',U,V \cup \{v\})$ by membership queries\;
    }
  }
  \Return{$\A_{\mathcal O}$}
  \caption{The learning algorithm for a session automaton $\A$}
  \label{table:algorithm}
\end{algorithm2e}

We now describe in detail our active learning algorithm for a given
session automaton $\A$ given in Table~\ref{table:algorithm}. It is
based on a loop which repeatedly constructs a closed table using
membership queries, builds the corresponding automaton and then asks
an equivalence query. This is repeated until $\A$ is learned.  An
important part of a active learning algorithm is the treatment of
counterexamples provided by the teacher as an answer to an equivalence
query.  Suppose that for a given $\A_{\mathcal O}$ constructed from a
closed table $\mathcal O=(T,U,V)$ the teacher answers by a
counterexample data word $w$.  Let $z=\snf(w)$.  If $z$ uses more
registers than available in the current alphabet, we extend the
alphabet and then the table. If the obtained table is not closed, we
restart from the beginning of the loop.  Otherwise -- and also if $z$
does not use more registers -- we use Rivest and Schapire's
\cite{RiSh:inference} technique to extend the table by adding a
suitable $v$ to $V$ making it non-closed.  The technique is based on
the notion of break-point that we now recall.  As $z$ is a
counterexample, (1) $z \in \symbL(\A_{\mathcal O}) \iff z \not\in
\symbL(\can{\A})$.  Let $z = z_1\cdots z_{m}$, with $z_i\in
\Sigma\times \Gamma$.  Then, for all $i$ with $1 \leq i \leq m+1$, let
$z$ be decomposed as $z = u_iv_i$ with $u_i, v_i\in(\Sigma\times
\Gamma)^*$, where $u_1 = v_{m+1} = \varepsilon$, $v_1 = u_{m+1} = z$
and the length of $u_i$ is equal to $i-1$ (we have also $z =
u_iz_iv_{i+1}$ for all $i$ such that $1 \leq i \leq m$).  Let $s_i\in
U$ be the state visited by $z$ just before reading the $i$th letter,
along the computation of $z$ on $\A_{\mathcal O}$: $i$ is a
break-point if $s_iz_iv_{i+1} \in \symbL(\A_{\mathcal O}) \iff
s_{i+1}v_{i+1} \notin \symbL(\can\A)$. Because of (1) such a
break-point must exist and can be obtained with $O(\log(m))$
membership queries by a binary search.  The word $v_{i+1}$ is called
the distinguishing word.  If $V$ is extended by $v_{i+1}$ the table is
not closed anymore ($row(s_i)$ and $row(s_iz_i)$ become
different). Now, the algorithm closes the table again, then asks
another equivalence query and so forth until termination. At each
iteration of the loop the number of rows (each of those correspond to
a state in the automaton $\can{\A}$) is increased by at least one.
Notice that the same counterexample might be given several times.  The
treatment of the counterexample only guarantees that the table will
contain one more row in its upper part.  We obtain the following:

\begin{thm}
  Let $\A$ be a session automaton over $\Sigma$ and $D$, using $k'$
  registers. Let $\can{\A}$ be the corresponding canonical session
  automaton. Let $N$ be its number of states, $k$ its number of
  registers and $M$ the length of the longest counterexample returned
  by an equivalence query. Then, the learning algorithm for $\A$
  terminates with at most $O(k |\Sigma| N^2 + N \log (M))$ membership
  and $O(N)$ equivalence queries.
\end{thm}
\proof
  This follows directly from the proof of correctness and complexity
  of Rivest and Schapire's algorithm \cite{BergR05,RiSh:inference}.
  Notice that the equivalence query cannot return a counterexample
  whose normal form uses more than $k$ registers, as such a word is
  rejected by both $\can{\A}$ (by definition) and by $\A_{\mathcal O}$
  (by construction). \qed

Let us discuss the complexity of our algorithm. In terms of the
canonical session automaton, the number of required membership and
equivalence queries is polynomial. When the session automaton $\A$ is
data deterministic, using the discussion after the proof of
Theorem~\ref{thm:saregular} over the size of $\can\A$, the overall
complexity of the learning algorithm is polynomial in the number of
states of $\A$, but exponential in the number of registers it uses
(with constant base). As usual, we have to add one exponent when we
consider session automata which are not data deterministic. In
\cite{HowarSJC12}, the number of equivalence queries is polynomial in
the size of the underlying automaton. In contrast, the number of
membership queries contains a factor $n^k$ where $n$ is the number of
states and $k$ the number of registers.  This may be seen as a
drawback, as $n$ is typically large.  Note that \cite{HowarSJC12}
restrict to deterministic automata, since classical register automata
are not determinizable.

\begin{figure}[ht]
\begin{center}
{\footnotesize
\noindent
\begin{tabular}{lllllllllllllll}
$
\begin{array}[t]{r||c}
\mathcal O_1& ~~\varepsilon~~ \\\hline\hline
\varepsilon&+\\
(b,\rreg{1})&-\\
\hline\hline
(a,\gfresh{1})&+\\
(b,\gfresh{1})\_ &-\\
\end{array}
$
&\raisebox{-2ex}{$\Rightarrow$}&
$
\begin{array}[t]{r||c||c}
\mathcal O_2&~~ \varepsilon~~ & (b,\rreg{1})\\\hline\hline
\varepsilon&+&-\\
(b,\rreg{1})&-&-\\
(a,\gfresh{1})&+&+\\
\hline\hline
(b,\rreg{1})\_ &-&-\\
(a,\gfresh{1})(a,\gfresh{1})&+&+\\
(a,\gfresh{1})(b,\rreg{1})&+&+\\
\end{array}
$
&\raisebox{-2ex}{$\Rightarrow$}&
$
\begin{array}[t]{r||c||c}
\mathcal O_3& ~~\varepsilon~~ & (b,\rreg{1})\\\hline\hline
\varepsilon&+&-\\
(b,\rreg{1})&-&-\\
(a,\gfresh{1})&+&+\\
\hline\hline
(a,\gfresh{2})&-&-\\
(b,\rreg{2})&-&-\\
(b,\rreg{1})\_ &-&-\\
(a,\gfresh{1})(a,\gfresh{1})&+&+\\
(a,\gfresh{1})(b,\rreg{1})&+&+\\
(a,\gfresh{1})(a,\gfresh{2})&-&+\\
(a,\gfresh{1})(b,\rreg{2})&-&-\\
\end{array}
$
&\raisebox{-2ex}{$\Rightarrow$}&
\end{tabular}

\vspace*{2ex}

\begin{tabular}{llllllllllll}
$
\begin{array}[t]{r||c||c}
\mathcal O_4~& ~\varepsilon~ & (b,\rreg{1})\\\hline\hline
\varepsilon&+&-\\
(b,\rreg{1})&-&-\\
(a,\gfresh{1})&+&+\\
(a,\gfresh{1})(a,\gfresh{2})&-&+\\
\hline\hline
(a,\gfresh{2})&-&-\\
(b,\rreg{2})&-&-\\
(b,\rreg{1})\_ &-&-\\
(a,\gfresh{1})(a,\gfresh{1})&+&+\\
(a,\gfresh{1})(b,\rreg{1})&+&+\\
(a,\gfresh{1})(b,\rreg{2})&-&-\\
(a,\gfresh{1})(a,\gfresh{2})(a,\gfresh{1})&-&-\\
(a,\gfresh{1})(a,\gfresh{2})(b,\rreg{1})&+&+\\
(a,\gfresh{1})(a,\gfresh{2})(a,\gfresh{2})&-&+\\
(a,\gfresh{1})(a,\gfresh{2})(b,\rreg{2})&-&+\\
\end{array}
$
&\raisebox{-2ex}{$\Rightarrow$}&
$
\begin{array}[t]{r||c||c||c}
\mathcal O_5& ~~\varepsilon~ & (b,\rreg{1})&(b,\rreg{2})\\\hline\hline
\varepsilon&+&-&-\\
(b,\rreg{1})&-&-&-\\
(a,\gfresh{1})&+&+&-\\
(a,\gfresh{1})(a,\gfresh{2})&-&+&-\\
(a,\gfresh{1})(a,\gfresh{2})(b,\rreg{1})&+&+&+\\
\hline\hline
(a,\gfresh{2})&-&-&-\\
(b,\rreg{2})&-&-&-\\
(b,\rreg{1})\_ &-&-&-\\
(a,\gfresh{1})(a,\gfresh{1})&+&+&-\\
(a,\gfresh{1})(b,\rreg{1})&+&+&-\\
(a,\gfresh{1})(b,\rreg{2})&-&-&-\\
(a,\gfresh{1})(a,\gfresh{2})(a,\gfresh{1})&-&-&-\\
(a,\gfresh{1})(a,\gfresh{2})(a,\gfresh{2})&-&+&-\\
(a,\gfresh{1})(a,\gfresh{2})(b,\rreg{2})&-&+&-\\
(a,\gfresh{1})(a,\gfresh{2})(b,\rreg{1})(a,\gfresh{1})&+&+&+\\
(a,\gfresh{1})(a,\gfresh{2})(b,\rreg{1})(b,\rreg{1})&+&+&+\\
(a,\gfresh{1})(a,\gfresh{2})(b,\rreg{1})(a,\gfresh{2})&-&+&-\\
(a,\gfresh{1})(a,\gfresh{2})(b,\rreg{1})(b,\rreg{2})&+&+&+\\
\end{array}
$
\end{tabular}}
\end{center}
\caption{\label{fig:learning}The successive observation tables}
\end{figure}

  \begin{gpicture}[name=hypo1,ignore]
    \unitlength=4 \gasset{Nw=4,Nh=4,Nmr=4,loopdiam=4}
    \gasset{AHangle=30,ilength=3.5,flength=3.5}
    \node[Nmarks=if,iangle=180,fangle=-90](0)(0,0){}
    \drawloop[loopangle=90](0){$a,\gfresh{1}$}
  \end{gpicture}
  \begin{gpicture}[name=hypo2,ignore]
    \unitlength=4
    \gasset{Nw=4,Nh=4,Nmr=4,loopdiam=4}
    \gasset{AHangle=30,ilength=3.5,flength=3.5}
    \node[Nmarks=if,iangle=180,fangle=-90](0)(0,0){}
    \node[Nmarks=f,fangle=0](1)(15,0){}
    \drawedge(0,1){$a,\gfresh{1}$}
    \drawloop[loopangle=90](1){$\begin{array}{l}
        a,\gfresh{1} \\ b,\rreg{1}
      \end{array}$}
  \end{gpicture}
  \begin{gpicture}[name=hypo3,ignore]
    \unitlength=4
    \gasset{Nw=4,Nh=4,Nmr=4,loopdiam=4}
    \gasset{AHangle=30,ilength=3.5,flength=3.5}
    \node[Nmarks=if,iangle=180,fangle=-90](0)(0,0){}
    \node[Nmarks=f,fangle=-90](1)(15,0){}
    \node(3)(30,0){}
    \drawedge(0,1){$a,\gfresh{1}$}
    \drawloop[loopangle=90](1){$\begin{array}{l}
        a,\gfresh{1} \\ b,\rreg{1}
      \end{array}$}
    \drawedge[curvedepth=2](1,3){$a,\gfresh{2}$}
    \drawloop[loopangle=90](3){$\begin{array}{l}
        a,\gfresh{2} \\ b,\rreg{2}
      \end{array}$}
    \drawedge[curvedepth=2](3,1){$b,\rreg{1}$}
  \end{gpicture}
  \begin{gpicture}[name=hypo4,ignore]
    \unitlength=4 \gasset{Nw=4,Nh=4,Nmr=4,loopdiam=4}
    \gasset{AHangle=30,ilength=3.5,flength=3.5}
    \node[Nmarks=if,iangle=180,fangle=-90](0)(0,0){}
    \node[Nmarks=f,fangle=-90](1)(15,0){} 
    \node(3)(30,0){}
    \node[Nmarks=f,fangle=-90](5)(45,0){}
    \drawedge(0,1){$a,\gfresh{1}$}
    \drawloop[loopangle=90](1){$\begin{array}{l} a,\gfresh{1} \\ b,\rreg{1}
      \end{array}$}
    \drawedge(1,3){$a,\gfresh{2}$}
    \drawloop[loopangle=90](3){$\begin{array}{l}
        a,\gfresh{2} \\ b,\rreg{2}
      \end{array}$}
    \drawedge[curvedepth=2](3,5){$b,\rreg{1}$}
    \drawedge[curvedepth=2](5,3){$a,\gfresh{2}$}
    \drawloop[loopangle=0](5){$\begin{array}{l}
        a,\gfresh{1} \\ b,\rreg{1} \\ b,\rreg{2}
      \end{array}$}
  \end{gpicture}
  \begin{figure}[t]
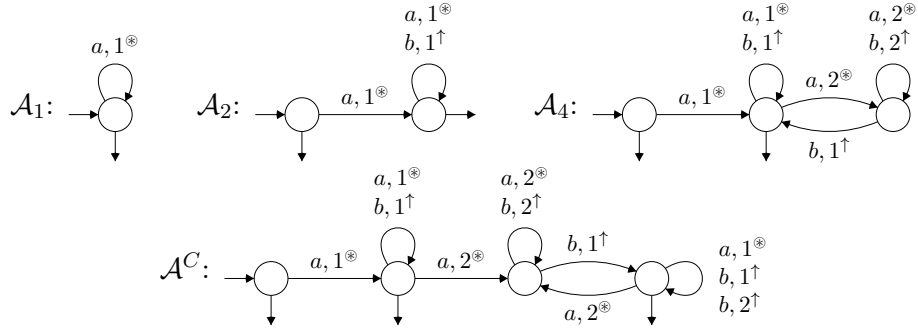

    \centering
    \begin{tabular}{c@{\hspace{2em}}c@{\hspace{2em}}c}
      $\A_1$: \raisebox{-1.7em}{\scalebox{.8}{\gusepicture{hypo1}}} & 
      $\A_2$: \raisebox{-1.7em}{\scalebox{.8}{\gusepicture{hypo2}}} & 
      $\A_4$: \raisebox{-1.7em}{\scalebox{.8}{\gusepicture{hypo3}}}
    \end{tabular}\\
    $\can \A$: \raisebox{-1.7em}{\scalebox{.8}{\gusepicture{hypo4}}}
    \caption{\label{fig:hypos}The successive hypothesis automata}
  \end{figure}  

\begin{exa}
  We apply our learning algorithm on the data language given by the
  automaton $\A$ of
  \figurename~\ref{fig:exampleCanonical}(a).  In
  \figurename~\ref{fig:learning} the successive observation tables
  constructed by the algorithm are given. To save space some letters
  whose rows contain only $-$'s are omitted.  In
  \figurename~\ref{fig:hypos} the successive automata constructed from
  the closed observation tables are given.  For sake of clarity we
  omit the sink states.  We start with the alphabet $\Sigma \times
  \Gamma_{1}= \{(a,\gfresh{1}),(a,\rreg{1}),(b,\gfresh{1}),
  (b,\rreg{1})\}$. We omit letters $(a,\rreg{1})$ and
  $(b,\gfresh{1})$.  Table $\mathcal O_1$ is obtained after
  initialization and closing by adding $(b,\rreg{1})$ to the top.  We
  use $\_$ to indicate that all letters will lead to the same row.
  From $\mathcal O_1$ the first hypothesis automaton $\A_1$ is
  constructed.  We suppose that the equivalence query gives back as
  counterexample the data word $(a,3)(b,3)$ whose normal form is
  $(a,\gfresh{1})(b,\rreg{1})$.  Here the break-point yields the
  distinguishing word $(b,\rreg{1})$.  We add it to $V$. The obtained
  table is not closed anymore. We close it by adding $(a,\gfresh{1})$
  to the top and get table $\mathcal O_2$ yielding hypothesis
  automaton $\A_2$. Notice that $\symbL(\A_2) = \symbL(\can\A) \cap
  (\Sigma \times \Gamma_{1})^*$.  This means that the equivalence
  query must give back a data word whose normal form is using at least
  $2$ registers (here $(a,7)(a,4)(b,7)$ with normal form
  $(a,\gfresh{1})(a,\gfresh{2})(b,\rreg{1})$).  As the word uses $2$
  registers, we extend the alphabet to $\Sigma \times \Gamma_{2}$ and
  obtain table $\mathcal O_3$.  We close the table and get $\mathcal
  O_4$.  From there we obtain the hypothesis automaton $\A_4$.  After
  the equivalence query we get
  $(a,\gfresh{1})(a,\gfresh{2})(b,\rreg{1})(b,\rreg{2})$ as normal
  form of the data word counterexample $(a,9)(a,3)(b,9)(b,3)$. After
  adding $(b,\rreg{2})$ to $V$ and closing the table by moving
  $(a,\gfresh{1})(a,\gfresh{2})(b,\rreg{1})$ to the top we get finally
  the table $\mathcal O_5$ from which the canonical automaton
  $\can{\A}$ is obtained and the equivalence query succeeds.
\end{exa}

\section{Conclusion}

In this paper, we developed a theory of session automata, which
form a robust class of data languages. In particular, they are closed under
union, intersection, and resource-sensitive complementation. Moreover, they
enjoy logical characterizations in terms of (a fragment of) MSO logic with
a predicate to compare data values for equality. Finally, unlike most other
automata models for data words, session automata have a decidable inclusion
problem. This makes them attractive for verification and learning.
In fact, we provided a complete framework for algorithmic learning of
session automata, making use of their canonical normal form. An
interesting direction to follow would be to try to apply those methods
to other models of automata dealing with data values like data
automata \cite{BojanczykDMSS11,Bjorklund10} or variable automata
\cite{DBLP:conf/lata/GrumbergKS10}.  As a next step, we plan to employ
our setting for various verification tasks. In particular, the next
step is to implement our framework, using possibly other learning
algorithms than the one of Rivest and Shapire that we presented in
this article, for instance using the LearnLib platform \cite{MRSL07}
or libalf \cite{BKKLNP10}.

\subsubsection*{Acknowledgments.} We are grateful to Thomas Schwentick
for suggesting the symbolic normal form of data words, and to the reviewers
for their valuable comments.


\begin{thebibliography}{10}

\bibitem{DBLP:conf/fm/AartsHKOV12}
F.~Aarts, F.~Heidarian, H.~Kuppens, P.~Olsen, and F.~W. Vaandrager.
\newblock Automata learning through counterexample guided abstraction
  refinement.
\newblock In {\em FM}, volume 7436 of {\em Lecture Notes in Computer Science},
  pages 10--27. Springer, 2012.

\bibitem{Angluin:regset}
D.~Angluin.
\newblock Learning regular sets from queries and counterexamples.
\newblock {\em Information and Computation}, 75(2):87--106, 1987.

\bibitem{BergGJLRS05}
T.~Berg, O.~Grinchtein, B.~Jonsson, M.~Leucker, H.~Raffelt, and B.~Steffen.
\newblock On the correspondence between conformance testing and regular
  inference.
\newblock In {\em FASE}, volume 3442 of {\em Lecture Notes in Computer
  Science}, pages 175--189. Springer, 2005.

\bibitem{BergR05}
T.~Berg and H.~Raffelt.
\newblock Model checking.
\newblock In {\em Model-based Testing of Reactive Systems}, volume 3472 of {\em
  Lecture Notes in Computer Science}. Springer, 2005.

\bibitem{Bjorklund10}
H.~Bj{\"o}rklund and {\relax Th}.~Schwentick.
\newblock On notions of regularity for data languages.
\newblock {\em Theoretical Computer Science}, 411(4-5):702--715, 2010.

\bibitem{BojanczykDMSS11}
M.~Bojanczyk, C.~David, A.~Muscholl, T.~Schwentick, and L.~Segoufin.
\newblock Two-variable logic on data words.
\newblock {\em ACM Trans. Comput. Log.}, 12(4):27, 2011.

\bibitem{BL2010}
M.~Boja{\'n}czyk and S.~Lasota.
\newblock An extension of data automata that captures {XPath}.
\newblock In {\em {LICS 2010}}, pages 243--252. IEEE Computer Society, 2010.

\bibitem{BCGK-fossacs12}
B.~Bollig, A.~Cyriac, P.~Gastin, and K.~Narayan~Kumar.
\newblock Model checking languages of data words.
\newblock In L.~Birkedal, editor, {\em {P}roceedings of {FoSSaCS}'12}, volume
  7213 of {\em Lecture Notes in Computer Science}, pages 391--405. Springer,
  2012.

\bibitem{BHLM-dlt2013}
B.~Bollig, P.~Habermehl, M.~Leucker, and B.~Monmege.
\newblock A~fresh approach to learning register automata.
\newblock In {\em {P}roceedings of the 17th {I}nternational {C}onference on
  {D}evelopments in {L}anguage {T}heory ({DLT}'13)}, volume 7907 of {\em
  Lecture Notes in Computer Science}, pages 118--130. Springer, 2013.

\bibitem{BKKLNP10}
B.~Bollig, J.-P. Katoen, C.~Kern, M.~Leucker, D.~Neider, and D.~Piegdon.
\newblock {libalf}: the automata learning framework.
\newblock In {\em CAV}, volume 6174 of {\em Lecture Notes in Computer Science},
  pages 360--364. Springer, 2010.

\bibitem{CobleighGP03}
J.~M. Cobleigh, D.~Giannakopoulou, and C.~S. Pasareanu.
\newblock Learning assumptions for compositional verification.
\newblock In {\em TACAS}, volume 2619 of {\em Lecture Notes in Computer
  Science}, pages 331--346. Springer, 2003.

\bibitem{Colcombet2011}
T.~Colcombet, C.~Ley, and G.~Puppis.
\newblock On the use of guards for logics with data.
\newblock In {\em Proceedings of MFCS'11}, volume 6907 of {\em Lecture Notes in
  Computer Science}, pages 243--255. Springer Berlin / Heidelberg, 2011.

\bibitem{Hig10}
C.~de~la Higuera.
\newblock {\em Grammatical Inference. Learning Automata and Grammars}.
\newblock Cambridge University Press, 2010.

\bibitem{DL-tocl08}
S.~Demri and R.~Lazi{\'c}.
\newblock {LTL} with the freeze quantifier and register automata.
\newblock {\em ACM Transactions on Computational Logic}, 10(3), 2009.

\bibitem{DBLP:conf/sigsoft/GiannakopoulouM03}
D.~Giannakopoulou and J.~Magee.
\newblock Fluent model checking for event-based systems.
\newblock In {\em ESEC / SIGSOFT FSE}, pages 257--266. ACM, 2003.

\bibitem{DBLP:conf/lata/GrumbergKS10}
O.~Grumberg, O.~Kupferman, and S.~Sheinvald.
\newblock Variable automata over infinite alphabets.
\newblock In {\em LATA}, volume 6031 of {\em Lecture Notes in Computer
  Science}, pages 561--572. Springer, 2010.

\bibitem{DBLP:conf/atva/GrumbergKS13}
O.~Grumberg, O.~Kupferman, and S.~Sheinvald.
\newblock An automata-theoretic approach to reasoning about parameterized
  systems and specifications.
\newblock In {\em ATVA}, volume 8172 of {\em Lecture Notes in Computer
  Science}, pages 397--411. Springer, 2013.

\bibitem{HV-infinity04}
P.~Habermehl and T.~Vojnar.
\newblock Regular model checking using inference of regular languages.
\newblock {\em Electronic Notes in Theoretical Computer Science},
  138(3):21--36, 2005.

\bibitem{HowarSJC12}
F.~Howar, B.~Steffen, B.~Jonsson, and S.~Cassel.
\newblock Inferring canonical register automata.
\newblock In {\em VMCAI}, volume 7148 of {\em Lecture Notes in Computer
  Science}, pages 251--266. Springer, 2012.

\bibitem{DBLP:conf/sfm/Jonsson11}
B.~Jonsson.
\newblock Learning of automata models extended with data.
\newblock In {\em SFM}, volume 6659 of {\em Lecture Notes in Computer Science},
  pages 327--349. Springer, 2011.

\bibitem{Kaminski1994}
M.~Kaminski and N.~Francez.
\newblock Finite-memory automata.
\newblock {\em Theoretical Computer Science}, 134(2):329--363, 1994.

\bibitem{KamTan06}
M.~Kaminski and T.~Tan.
\newblock Regular expressions for languages over infinite alphabets.
\newblock {\em Fundamenta Informaticae}, 69(3):301--318, 2006.

\bibitem{KaminskiZ10}
M.~Kaminski and D.~Zeitlin.
\newblock Finite-memory automata with non-deterministic reassignment.
\newblock {\em International Journal of Foundations of Computer Science},
  21(5):741--760, 2010.

\bibitem{DBLP:conf/ccs/KurtzKW07}
K.~O. K{\"u}rtz, R.~K{\"u}sters, and T.~Wilke.
\newblock Selecting theories and nonce generation for recursive protocols.
\newblock In P.~Ning, V.~Atluri, V.~D. Gligor, and H.~Mantel, editors, {\em
  FMSE}, pages 61--70. ACM, 2007.

\bibitem{KST2012}
A.~Kurz, T.~Suzuki, and E.~Tuosto.
\newblock On nominal regular languages with binders.
\newblock In L.~Birkedal, editor, {\em {P}roceedings of {FoSSaCS}'12}, volume
  7213 of {\em Lecture Notes in Computer Science}, pages 255--269. Springer,
  2012.

\bibitem{DBLP:conf/fmco/Leucker07}
M.~Leucker.
\newblock Learning meets verification.
\newblock In {\em FMCO}, volume 4709 of {\em Lecture Notes in Computer
  Science}, pages 127--151. Springer, 2007.

\bibitem{MRSL07}
T.~Margaria, H.~Raffelt, B.~Steffen, and M.~Leucker.
\newblock The {LearnLib} in {FMICS-jETI}.
\newblock In {\em ICECCS}, pages 340--352. IEEE Computer Society Press, 2007.

\bibitem{MPW92}
R.~Milner, J.~Parrow, and D.~Walker.
\newblock A calculus of mobile processes, {P}arts {I} and {II}.
\newblock {\em Information and Computation}, 100:1--77, Sept. 1992.

\bibitem{Neven2004}
F.~Neven, {\relax Th}.~Schwentick, and V.~Vianu.
\newblock Finite state machines for strings over infinite alphabets.
\newblock {\em ACM Transactions on Computational Logic}, 5(3):403--435, 2004.

\bibitem{RiSh:inference}
R.~Rivest and R.~Schapire.
\newblock Inference of finite automata using homing sequences.
\newblock {\em Information and Computation}, 103:299--347, 1993.

\bibitem{SakIke00}
H.~Sakamoto and D.~Ikeda.
\newblock Intractability of decision problems for finite-memory automata.
\newblock {\em Theoretical Computer Science}, 231:297--308, 2000.

\bibitem{Segoufin06}
L.~Segoufin.
\newblock Automata and logics for words and trees over an infinite alphabet.
\newblock In Z.~{\'E}sik, editor, {\em CSL 2006}, volume 4207 of {\em LNCS},
  pages 41--57. Springer, 2006.

\bibitem{DBLP:conf/popl/Tzevelekos11}
N.~Tzevelekos.
\newblock Fresh-register automata.
\newblock In T.~Ball and M.~Sagiv, editors, {\em POPL}, pages 295--306. ACM,
  2011.

\end{thebibliography}

\def\Nst#1{$#1^{st}$}\def\Nnd#1{$#1^{nd}$}\def\Nrd#1{$#1^{rd}$}\def\Nth#1{$#1^{th}$}

\end{document}